\documentclass[ejs]{imsart}
\usepackage{imsart}

\usepackage{amsthm,amsmath}
\RequirePackage[numbers]{natbib} 
\RequirePackage[colorlinks,citecolor=blue,urlcolor=blue]{hyperref}

\usepackage{mathtools, mathrsfs} 
\usepackage{bbm} 
\usepackage[psamsfonts]{amssymb}
\usepackage{xcolor, color}
\usepackage{multirow}
\usepackage[caption = false]{subfig}
\usepackage{enumerate}

\usepackage[linesnumbered, ruled]{algorithm2e}
\SetKwRepeat{Do}{do}{while}%

\doi{10.1214/154957804100000000}
\pubyear{0000}
\volume{0}
\firstpage{1}
\lastpage{1}

\startlocaldefs

\newtheorem{thm}{Theorem}[section]
\newtheorem{dfn*}{Definition}[section]
\newtheorem{prop}{Proposition}[section]

\newtheorem{ass}{Assumption} 
\newtheorem*{assumption*}{\assumptionnumber}
\providecommand{\assumptionnumber}{}


\DeclareMathOperator*{\argmin}{\arg\!\min}

\newcommand\independenT{\protect\mathpalette{\protect\independenT}{\perp}}
\def\independenT#1{\mathrel{\rlap{$#1$}\mkern2mu{#1}}}

\newcommand\smtop{\mkern-2mu\raise.25ex\hbox{$\scriptscriptstyle\top$}\mkern-3mu}

\def\bs{\boldsymbol}

\def\a{\alpha}
\def\b{\beta}
\def\g{\gamma}
\def\th{\theta}

\def\t{\tau}

\def\d{\delta}

\def\vep{\varepsilon}

\def\i{\mathbbm{1}}

\def\om{\Omega}

\def\R{\mathbb{R}}

\def\SS{\mathscr{S}}

\def\indep{\independenT{\perp}}

\def\vecv{\textnormal{vec}}

\endlocaldefs

\begin{document}

\begin{frontmatter}

\title{On Forward Sufficient Dimension Reduction for Categorical and Ordinal Responses}
\runtitle{On Forward Sufficient Dimension Reduction}


\author{\fnms{Harris} \snm{Quach}\ead[label=e1]{hxq5@psu.edu}}
\and
\author{\fnms{Bing} \snm{Li}\ead[label=e2]{bxl9@psu.edu}}

\address{The Pennsylvania State University,
	University Park,
	USA.\\
	\printead{e1,e2}}

\runauthor{Quach and  Li}

\begin{abstract}
	We introduce a forward sufficient dimension reduction method for categorical or ordinal responses by extending the outer product of gradients and minimum average variance estimator to categorical and ordinal-categorical generalized linear models. Previous works in this direction extend forward regression to binary responses, and are applied in a pairwise manner for multi-category data, which is less efficient than our approach. Like other forward regression-based sufficient dimension reduction methods, our approach avoids the relatively stringent distributional requirements necessary for inverse regression alternatives. We show the consistency of our proposed estimator and derive its convergence rate. We develop an algorithm for our methods based on repeated applications of available algorithms for forward regression. We also propose a clustering-based tuning procedure to estimate the bandwidth. The effectiveness of our estimator and related algorithms is demonstrated via simulations and applications.          
\end{abstract}





\end{frontmatter}


\section{Introduction}

The Outer Product of Gradients and the Minimum Average Variance Estimator of \citet{xia2002adaptive} are two popular methods for sufficient dimension reduction \citep{li1991sliced, cook1991comment,li2018sufficient} because of the weak distributional assumptions they require{,} and their efficient and stable performance.
However, both {methods} are based on the derivative of the response variable, and therefore are inappropriate for categorical responses. 
The purpose of this paper is to extend these methods to the categorical and ordinal categorical responses by imposing a multivariate link function on the conditional mean of the response in a {localized multivariate} generalized linear model.

Let $Y$ denote {a} response variable and $X$ a $p$-dimensional predictor. 
Sufficient dimension reduction (SDR) estimates a lower dimensional function of $X$ that retains all the relevant information about $Y$ that is available in $X$. That is, we seek to estimate a linear function of $X$, represented by $\beta^{\smtop} X$, where $\beta \in \R^{p \times d}$ with $d < p$, such that
\begin{align}
\label{ass:sdr}
Y \indep X | \beta^{\smtop} X
{,}
\end{align}
{where $A \indep B | C$ indicates conditional independence.}
%

Since relation \eqref{ass:sdr} is identifiable only up to the column space of $\beta$, the parameter of interest is the subspace $\mathrm{span}(\beta)$. For any $\beta$ {satisfying} \eqref{ass:sdr}, the corresponding {$\mathrm{span}(\beta)$} is referred to as {an} SDR subspace, and the columns of $\beta$ are referred to as {the} {sufficient} directions. While existence of some SDR subspace {is evident by} taking $\beta$ as the identity, existence of a minimal SDR subspace requires some mild conditions that are given in generality by \cite{yin2008successive}. We will assume existence of the minimal SDR subspace, {refer} to {it} as the \textit{Central Subspace (CS)} and {denote it} by $\SS_{Y|X}$. The subspace $\SS_{Y|X}$ is minimal in the sense that for any $\beta$ such that \eqref{ass:sdr} holds, {$\SS_{Y|X} \subseteq \mathrm{span}(\beta)$}. The goal of SDR, in the context of \eqref{ass:sdr}, is to find $\beta$ such that $\mathrm{span}(\beta) = \SS_{Y|X}$.       

In many statistical applications, the regression function, $E(Y|X)$, is of primary interest. {For this situation, \citet{cook2002dimension} formulated a weaker form of SDR by requiring}
\begin{align}
\label{ass:sdr-cms}
Y \indep E(Y|X) | \b^{\smtop} X
\end{align}
{for some matrix $\b$ with fewer columns than rows. This problem is called SDR for the conditional mean.}
{As shown in \citet{cook2002dimension},} \eqref{ass:sdr-cms} is equivalent to $E(Y|X) = E(Y|\b^{\smtop}X)$. The minimal SDR subspace such that \eqref{ass:sdr-cms} holds {is called} the \textit{Central Mean Subspace (CMS)} and is denoted by $\SS_{E(Y|X)}$. The CMS for $E(Y|X)$ is analogous to the CS for $Y|X$, and so $\SS_{E(Y|X)}$ exists under mild conditions as well. We will assume that $\SS_{E(Y|X)}$ exists, and the aim of SDR for {the conditional mean} is to find $\b$ such that $\mathrm{span}(\b) = \SS_{E(Y|X)}$.      
{ The mentioned OPG and MAVE are, in fact, both estimators of the central mean subspace.} 


{
	In this paper we extend OPG and MAVE to situations where the response is categorical or ordinal-categorical variables. These types of responses are very common in practice. In particular, ordinal-categorical responses are one of the prevalent data forms in market analysis (see, for example, \citet{zhang2021generalized}).
	We employ a localized multivariate generalized linear model (GLM), where the categorical/ordinal  responses are modeled through multivariate link functions. The central subspace, which in this case coincides with the central mean subspace, is estimated by the gradient of the canonical parameter of the exponential family that generates the GLM. Through the use of the multivariate link function, we avoid estimating the derivative of a discrete variable, which is essentially what one would do if one applies OPG or MAVE directly to this setting. 
	
	A direct precursor of our method is \citet{adragni2018minimum}, which introduced the Minimum Average Deviance Estimator (MADE). This method, however, was designed for binary classification and treats a multi-label problem as pairs of binary problems. In comparison, our approach treats all classes simultaneously and achieves higher efficiency by doing so. Another related work is \citet{lambert2006local}, which proposed  a method called GSIM. While this method also employs a localized multivariate GLM to perform SDR, it is more akin to the Average Derivative Estimator (ADE, \citet{hardle1989investigating}) than OPG, and cannot estimate the central subspace exhaustively. A more detailed description of its difference from the current approach will be given in section  \ref{sec:opcg}, as it requires  technical notations unsuitable for an introduction.}
We also introduce a clustering-based tuning procedure for {the proposed methods, which} avoids the need to select a prediction method for conventional cross-validation.


{
	The rest of the paper is organized as follows. In section \ref{sec:glm_opcg}{, we} extend OPG and MAVE using a localized multivariate generalized linear model for sufficient dimension reduction. This allows us to incorporate a multivariate link function for SDR. This framework is more general than the context of categorical and ordinal data. 
	In section \ref{sec:link}, we describe the specific forms of the localized multivariate GLM for categorical and ordinal response using commonly used multivariate link functions. 
	In section \ref{sec:fisher}, we establish the Fisher consistency of our method, and in section \ref{sec:consis}, we prove the consistency and develop the convergence rate. 
	In section \ref{sec:imp}, we further discuss issues involved in implementation, such as order determination and bandwidth selection.
	In section \ref{sec:examples}, we provide some simulation comparisons with existing methods and data applications, which involve both purely categorical and ordinal categorical data{.}
	We conclude in section \ref{sec:con} with a discussion. All proofs are relegated to the online {S}upplementary {M}aterials.  
}

\section{ {Local Likelihood} Functions {and Two Estimators} }\label{sec:glm_opcg}

{
	\subsection{Multivariate generalized linear and nonlinear models}
	  
	While the primary goal of this paper is to address the practical issue of how to deal categorical or ordinal categorical responses in forward SDR, it is easier to explain the basic idea in terms of a generic multivariate generalized linear model and its localization. Let $X$ and $Y$ be random vectors defined on $\Omega_X \subseteq \R ^ p$ and $\Omega_Y \subseteq \R ^ m$, respectively. They follow a multivariate generalized linear model \citep{kim2006univariate} if the conditional density of $Y$ given $X$ with respect to a measure $\nu$ on $\Omega_Y$ is
	\begin{align}\label{eq:multivariate GLM}
	f_{Y|X}(y|x) = \exp \{ \theta (a + B ^{\smtop} x) ^{\smtop} y - b ( \theta (a +  B ^{\smtop} x )) \},
	\end{align}
	where $a \in \R^m$ and $B \in \R ^ {p \times m}$ are  the regression parameters, and the function $\theta: \R^m \to \R^m$ is the canonical parameter. The function $b:\R^m \to \R$ is related to the conditional mean $E(Y|X)$ by
	$E(Y|X = x) = \dot b \{ \theta (a + B ^{\smtop} x) \}$, where $\dot b:\R^m \to \R^m $ is the gradient of the function $b$, which is injective if $\mathrm{Var}(Y|X)$ is positive definite. 
	%

	Adopting the GLM terminology in \cite{mccullagh1989generalized}, let $\t = \t(\th)$ be the mean parameter defined by $\t(\th) = E(Y|X) = \dot b(\th)$, let $\eta = a + B^{\smtop}x$ be the linear predictor, and let $\psi$ be the link function defined by $\psi(\t) = \eta$. Then $\theta = \dot b^{-1} (\t) = \dot b^{-1} \circ \psi^{-1} (\eta) = \t^{-1} \circ \psi^{-1} (\eta)$, and the conditional density in \eqref{eq:multivariate GLM} becomes
	\begin{align}\label{eqn:noncan density}
	f_{Y|X}(y|x) = \exp \{ (\t^{-1} \circ \psi^{-1}) (a + B^{\smtop}x ) ^{\smtop} y - 
	(b \circ \t^{-1} \circ \psi^{-1} ) ( a + B^{\smtop} x ) \}.
	\end{align}
	The canonical link function is defined so that $\th(\cdot)$ is the identity mapping; that is, $\psi = \dot b^{-1}$.
	Under the canonical link, the conditional density reduces to 
	\begin{align}\label{eqn:can density}
	f_{Y|X}(y|x) = \exp \{ (a + B^{\smtop}x ) ^{\smtop} y - 
	b  ( a + B^{\smtop} x ) \}.
	\end{align}

	The model underlying our SDR method is
	\begin{align}\label{eq:fully nonparametric}
	f_{Y|X}(y|x) = \exp [ \theta (x) ^{\smtop} y - b \{ \theta ( x ) \} ],
	\end{align}
	where $\theta (x)$ is an unknown nonlinear function. Thus, \eqref{eq:fully nonparametric} is a type of multivariate nonlinear regression, which we call the multivariate generalized nonlinear model (GNM), as $Y$ depends on $X$ nonlinearly through the canonical parameter of an exponential family. Our goal is to perform sufficient dimension reduction under this model. It can be shown that the density $f_{Y|X}(y|x)$ depends on $x$ only through the conditional mean $E(Y|X = x)$, which implies that the central subspace $ \SS_{Y|X}$ and the central mean subspace $\SS_{E(Y|X)}$ are the same \citep{cook2002dimension}. The  forward regression methods  such as OPG and MAVE  are natural candidates for extensions to target such situations. 
	
	\subsection{{Outer Product of Canonical Gradients}} \label{sec:opcg} 
	While model (\ref{eq:fully nonparametric}) is fully nonparametric, the way we estimate the function $\theta(\cdot)$ is through the localized multivariate GLM, in much the same way Xia, Tong, Li, and Zhu (2001) applies the local linear regression to estimate the central mean subspace in their problem. 
	Specifically, for a fixed member $x_0$ of $\Omega_X$, we introduce the local linear predictor
	\begin{align}
	\label{eqn:lin_pred}
	\eta_{x_0} (x) = a_{x_0} + B_{x_0} ^{\smtop} (x - x_0).
	\end{align}
	Under the link function $\psi$, we have the following localized multivariate GLM: 
	\begin{align*}
	f_{Y|X}(y|x) = \exp[ (\t^{-1} \circ \psi^{-1}) \{ \eta_{x_0}(x) \}^{\smtop} y - (b \circ \t^{-1} \circ \psi^{-1} )\{ \eta_{x_0}(x) \} ]
	,
	\end{align*}
	which closely resembles \eqref{eqn:noncan density}.
	Under the canonical link, this conditional density reduces to 
	\begin{align*}
	f_{Y|X}(y|x) = \exp[\eta_{x_0}(x)^{\smtop} y -  b  \{ \eta_{x_0}(x) \} ].
	\end{align*}
	
	Let $K: \R ^ p \to \R$ be a kernel function with bandwidth $h$. At the population level, our method amounts to minimizing the expectation of the negative local likelihood weighted by the kernel; that is, we minimize, over $a_{x_0} \in \R ^ m$ and $B_{x_0} \in \R ^ {p \times m}$, the following objective function:
	\begin{align*}
	E [  K \{ (X - x_0)/ h \} ( - [ (\t^{-1} \circ \psi^{-1} ) \{ \eta_{x_0}( x )\} ]^{\smtop} Y +  ( b \circ \t^{-1} \circ \psi^{-1} )  \{  \eta_{x_0}(x) \}  )  ].
	\end{align*}
	At the sample level, we minimize, for $x \in \Omega_X$, the objective function 
	\begin{align}
	\label{eqn:opcg-loc-like}
	\begin{split}
	\ell(x,a, B) = & n ^{-1} \sum_{i=1} ^ n  W_{ix}(h) ( -[ (\t^{-1} \circ \psi^{-1}  ) \{a + B^{\smtop} (X_i - x) \} ] ^{\smtop} Y_i  \\
	& + (b \circ \t^{-1} \circ \psi^{-1} ) \{a + B^{\smtop} (X_i  - x) \} ),
	\end{split}
	\end{align}
	where $W_{ix}(h) = K ( (X_i - x )/ h)$, 
	over $a \in \R ^ m$, $B  \in \R ^ {p \times m}$. Let $(\hat a (h,x), \hat B(h, x))$ be the minimizer of \eqref{eqn:opcg-loc-like}, and let $\hat a_i = \hat a (h, X_i)$, $\hat B_i = \hat B(h, X_i)$. Our proposed OPCG amounts to assembling  the minimizers $\hat B_1, \ldots, \hat B_n$ together by PCA to estimate the central mean subspace.
	That is, we use the first $d$ eigenvectors of the matrix $\hat \Lambda_{\mathrm{opcg}}= n^{-1}\sum_{i=1}^n \hat B_i \hat B_i^{\smtop}$ as an estimate of a basis of the central mean subspace. 
	We denote the $p \times d$ matrix formed by the first $d$ eigenvectors as $\hat \b_{\mathrm{opcg}}$.
	We summarize the general estimation procedure for OPCG in Algorithm \ref{alg:opcg}.
	In the following, a set of numbers, vectors, or matrices, $u_1, \ldots, u_n$, will be abbreviated at $u_{1:n}$.
	
	\begin{algorithm}[H] 
		\label{alg:opcg}
		\SetAlgoLined
		For $j=1,\ldots,n$, set $x=X_j$ and  estimate $\hat a_j, \hat B_j$ by minimizing \eqref{eqn:opcg-loc-like}, which can be performed, for example, using Algorithm \ref{alg:opcg_newton}\;
		Construct the matrix $\hat \Lambda_{\mathrm{opcg}}= n^{-1}\sum_{i=1}^n \hat B_i \hat B_i^{\smtop}$\;
		Estimate the $d$ leading eigenvectors of $\hat \Lambda_{\mathrm{opcg}}$, $\hat \eta_1,\ldots, \hat \eta_d$\; 
		Set $\hat \b_{ \mathrm{opcg} } = (\hat \eta_1 \cdots \hat \eta_d) \in \R^{p \times d}$.
		\caption{The Outer Product of Canonical Gradients Estimator}
	\end{algorithm}
	 
	{The minimization in step 1 can be performed by methods such as Newton-Raphson \citep{mccullagh1989generalized} or conjugate gradients \citep{dai2001efficient}. The Newton-Raphson method specialized to our context is outlined in the Supplementary Material, section 2. 
	}
	We carry out step 1 in Algorithm \ref{alg:opcg} by a Newton-Raphson algorithm that minimizes the full negative log-likelihood
	\begin{align*} 
	\ell( a_{1:n}, B_{1:n}; Y_{1:n}, X_{1:n})  
	= \sum_{j=1}^n \ell_j(a_j,B_j; Y_{1:n}, X_{1:n} )
	,
	\end{align*}
	where  
	\begin{align*} 
	\ell_j(a_j, B_j; Y_{1:n}, X_{1:n}) 
	& = n^{-1}  \sum_{i=1}^n  W_{ij}(h) ( - [ ( \t^{-1} \circ \psi^{-1} ) \{a_j + B_j^{\smtop} (X_i - X_j) \} ]^{\smtop}  Y_i \\
	& + ( b \circ \t^{-1} \circ \psi^{-1} )\{ a_j + B_j^{\smtop} (X_i - X_j) \}  ) 
	,
	\end{align*}
	with $W_{ij}(h) = K ( (X_i - X_j )/ h)$. 
	Let $A_j = (a_j, B_j^{\smtop})^{\smtop}  \in \R^{ (p+1) \times m }$,
	$c_j = \vecv(A_j^{\smtop} ) \in \R^{m(p+1)}$, and 
	$V_{ij} = (1 , (X_i - X_j)^{\smtop} )^{\smtop}  \in \R^{(p+1)}$. 
	Abbreviate $\ell_j(a_j, B_j; Y_{1:n}, X_{1:n})$ by $\ell_j(c_j)$, and denote the gradient vector and Hessian matrix of $\ell_j (c_j)$ by $S_j(c_j)$ and $J_j(c_j)$. Then, by straightforward computation,
	\begin{align*}
	\ell_{j}(c_j ) 
	& = \sum_{i=1}^n  W_{ij}(h) ( - [ ( \t^{-1} \circ \psi^{-1} ) \{(V_{ij}^{\smtop} \otimes I_m)  c_j  \} ]^{\smtop}  Y_i  \\
	& + ( b \circ \t^{-1} \circ \psi^{-1} ) \{ (V_{ij}^{\smtop} \otimes I_m) c_j  \}  ) 
	, \\
	S_j(c_j) 
	& = \sum_{i=1}^n  W_{ij} (h) (V_{ij} \otimes I_m)
	\bigg \{ -
	\bigg [ \frac{ \partial ( \t^{-1} \circ \psi^{-1} )  \{ (V_{ij} \otimes I_m)^{\smtop} c_j  \}   }
	{\partial \{ (V_{ij} \otimes I_m)^{\smtop} c_j  \} }  
	\bigg ]^{\smtop}  Y_i  \\
	& + 
	\bigg [ \frac{ \partial ( b \circ \t^{-1} \circ \psi^{-1} ) \{ (V_{ij} \otimes I_m)^{\smtop} c_j   \}   }
	{\partial \{ (V_{ij} \otimes I_m)^{\smtop} c_j  \} }  \bigg ]  
	\bigg \} ,\\ 
	J_j(c_j) 
	& = \sum_{i=1}^n  W_{ij} (h) (V_{ij} \otimes I_m)
	\bigg \{ - 
	\bigg [ \frac{ \partial^2 ( \t^{-1} \circ \psi^{-1} )  \{ (V_{ij} \otimes I_m)^{\smtop} c_j  \}   }{\partial \{ (V_{ij} \otimes I_m)^{\smtop} c_j  \}
		\partial \{ (V_{ij} \otimes I_m)^{\smtop} c_j  \}^{\smtop}  }  \bigg ]^{\smtop}  Y_i  \\
	& + 
	\bigg [ \frac{ \partial^2 ( b \circ \t^{-1} \circ \psi^{-1} ) \{ (V_{ij} \otimes I_m)^{\smtop} c_j  \}   }{\partial \{ (V_{ij} \otimes I_m)^{\smtop} c_j  \} \partial \{ (V_{ij} \otimes I_m)^{\smtop} c_j  \}^{\smtop} }  \bigg ]  
	\bigg \} (V_{ij} \otimes I_m)^{\smtop}  
	.
	\end{align*}
	Given an initial estimate for $c_j$, we iterate
	$ \hat c_j^{(r+1)} = \hat c_j^{(r)} + J_j^{-1} (\hat c_j^{(r)} ) S_j(\hat c_j^{(r)}) $
	until convergence with respect to some criteria, such as the relative Euclidean distance between iterations
	${ || \hat c_j^{(r+1)} - \hat c_j^{(r+1)} ||  } / { ||\hat c_j^{(r)} ||  } < \vep $, 
	for some chosen tolerance $\vep > 0$. 
	We denote the result after convergence as $\hat c_j$, and construct $\hat A_j = \mathrm{mat}(\hat c_j)^{\smtop} \in \R^{(p+1) \times m}$, where the $\mathrm{mat}$ operation takes the first $p+1$ entries of $\hat c_j $ as the first column, the next $p+1$ entries as the second column, etc. From $\hat A_j$, we extract the estimate $\hat B_j$ by removing the first row of $\hat A_j$.
	We then construct $\hat \Lambda_{\mathrm{opcg}} = n^{-1} \sum_{j=1}^n \hat B_j \hat B_j^{\smtop}$ and take the $d$ leading eigenvectors $\hat \eta_1,..., \hat \eta_d$ as our estimate for the SDR directions, $\hat \b_{\mathrm{opcg}} = (\hat \eta_1,..., \hat \eta_d)$.
	
	Because we have reformulated OPCG as fitting a GLM, we can construct the initial values for our procedure as in \cite{mccullagh1989generalized}. 
	Let 
	\begin{align*}
	\theta(Y) = ( \th(Y_1),\ldots, \th(Y_n))^{\smtop}, 
	\quad 
	V_j=(V_{1j}\ldots,V_{nj})^{\smtop}
	.
	\end{align*} 
	The initial value of $A_j$ is set as the least squares estimator $\hat A_j^{(0)} = (V_{ j}^{\smtop} V_{j})^{-1} V_{ j}^{\smtop}  \th(Y)$, from which we construct the initial estimate for $\hat c_j^{(0)}$ as $\vecv  \{ (\hat A_j^{(0)}  )^{\smtop} \} $. We summarize the Fisher-Scoring algorithm for $\hat V_j$ in Algorithm \ref{alg:opcg_newton}.
	
	\begin{algorithm}[H] 
		\label{alg:opcg_newton}
		\SetAlgoLined
		Set $\vep >0$ and construct $\hat c_j^{(0)} = \vecv(  (V_{ j}^{\smtop} V_{j})^{-1} V_{ j} \th(Y)^{\smtop} )$\;
		For $r=0,1,2,\ldots$, construct $\hat c_j^{(r+1)} = \hat c_j^{(r)} + J_j^{-1} (\hat c_j^{(r)} ) S_j(\hat c_j^{(r)}) $ until ${|| \hat c_j^{(r+1)} - \hat c_j^{(r)} ||  } / { ||\hat c_j^{(r)} ||  } < \vep $\;
		Set $\hat c_j$ as the most recent iteration, $\hat c_j = \hat c_j^{(r+1)}$\;
		Set $\hat A_j = \{ \mathrm{mat}(\hat c_j) \}^{\smtop}$. Then $\hat B_j$ is the matrix obtained by removing the first row of $\hat A_j$. 
		\caption{A Fisher-Scoring Algorithm for $\hat B_j$}
	\end{algorithm}
	
	In passing, we note that the word ``canonical'' in Outer Product of Canonical Gradients refers to the fact that we use the canonical gradient $\partial \theta (x)^{\smtop} / \partial x$ to estimate the central mean subspace. In particular, it does not imply that we must use the canonical link function. 
}

{
	\citet{lambert2006local} introduced a related method, called GSIM, that uses the local
	likelihood of a multivariate GLM to perform sufficient dimension reduction with categorical
	responses. However, there are two important differences.
}
{First,} GSIM targets the gradient of the conditional mean {rather than the canonical parameter.} 
{Second, t}heir proposed estimator for $\b$ is the average of the derivative estimates for the conditional mean and therefore is more similar to the Average Derivative Estimator {(ADE)} \citep{hardle1989investigating} than OPG {or MAVE}. The {ADE} requires the gradient to have {a nonzero} mean, fails when the distribution of $X$ is symmetric, and can only recover one SDR direction at a time \citep{xia2002adaptive}.
Furthermore, while GSIM { can be applied to multivariate $Y \in \R^m$, the procedure estimates one SDR direction for each component of the conditional mean by assuming $\mu(X) = \{ \mu^1(X),\ldots, \mu^m(X) \} = \{ \mu^1(\beta_1^{\smtop}X),\ldots, \mu^m(\beta_m^{\smtop}X) \}$. If we consider the canonical link, then GSIM estimates each SDR direction $\beta_k$ by the average of estimated partial derivatives, $\hat \b_k = n^{-1} \sum_{j=1}^n {\partial \hat \mu^k(X_j) / \partial x} \in \R^p$, for $k=1,\ldots,m$. This implies GSIM always returns $m$ SDR directions and the SDR directions do not incorporate information available from other components of the gradient.} 
{
	In comparison, our method targets the canonical parameter directly, fully recovers the central mean space and treats $Y$ jointly using a multivariate link function.
}
By forgoing the conditional mean, our estimation procedure is simpler relative to GSIM, { particularly when the canonical link function is used.}

{
	\subsection{{Multivariate} Minimum Average Deviance Estimator}
	Since $(X,Y)$ satisfies both the multivariate GNM \eqref{eq:fully nonparametric} and the SDR condition \eqref{ass:sdr} for some $\beta \in \R^{p \times d}$, for each $x_0 \in \Omega_X$, the matrix $B_{x_0}$ in the local linear predictor \eqref{eqn:lin_pred} must be of the form  $\beta C_{x_0}$ for some $C_{x_0} \in \R^{d \times m}$.
	The reparameterized negative log-likelihood is now
	\begin{align}
	\label{eqn:made-nll-full}
	\ell( a_{1:n}, C_{1:n}, \beta; Y_{1:n}, X_{1:n})  
	= \sum_{j=1}^n \ell_j(a_j,C_j, \beta)
	,
	\end{align}
	where $a_j$ and $C_j$ stand for $a_{x_j}$ and $C_{x_j}$, respectively, and  
	\begin{align}
	\label{eqn:made-nll-loc}
	\ell_j(a_j, C_j, \beta) 
	& = \sum_{i=1}^n  W_{ij}(h) [  \{a_j + C_j^{\smtop} \beta^{\smtop}(X_i - X_j) \}^{\smtop}  Y_i  - b( a_j + C_j^{\smtop} \beta^{\smtop} (X_i - X_j) )  ] ,
	\end{align}
	with $W_{ij}(h)$ being the normalized kernel weight $K  \{( X_i - X_j )/h\} / \sum_{i=1}^n  K  \{( X_i - X_j )/h\} $. Minimizing \eqref{eqn:made-nll-full} iteratively between $a_1,...,a_n, C_1,...,C_n$ and $\beta$ until convergence, we obtain the estimate of $\b$ directly, which we denote by $\hat \beta_{\mathrm{m-made}}$ and refer to it as the \textit{Multivariate Minimum Average Deviance Estimator (M-MADE)} for $\SS_{E(Y|X)}$. 
	Since $\beta$ in \eqref{eqn:made-nll-loc} is identifiable up to orthogonal transformations, the $\b$ solutions to the minimization of \eqref{eqn:made-nll-full} are not unique. 	
	We summarize the general estimation procedure for M-MADE in Algorithm \ref{alg:made}.
	A Fisher-Scoring algorithm for estimating $\beta$ in step 2 is provided in the Supplementary Materials. 
}

\begin{algorithm}[H]
	\label{alg:made}
	\SetAlgoLined 
	Hold $\beta$ fixed. For $j=1,\ldots,n$, set $x=X_j$ and estimate $\hat a_j$, $\hat C_j$ by minimizing \eqref{eqn:opcg-loc-like} with $B^{\smtop}(X_i - x)$ replaced with $C^{\smtop}\beta^{\smtop}(X_i - x)$\;
	Plug $\hat a_1,\ldots,\hat \a_n$ and $\hat C_1,\ldots,\hat C_n$ into \eqref{eqn:made-nll-full} and hold them fixed. Estimate $\b$ by minimizing \eqref{eqn:made-nll-full}\;
	Plug the minimizer $\beta$ from Step 2 into Step 1 and repeat Steps 1 and 2 until the estimates for $\b$ converge\;
	Set $\hat \b_{ \mathrm{made} }$ to be the most recent estimate of $\beta$ in Step 2.
	\caption{The Minimum Average Deviance Estimator}
\end{algorithm}
As in step 1 of Algorithm \ref{alg:opcg}, the minimization in step 3 can be performed by a Newton-Raphson method or conjugate gradients. The former is outlined in the Supplementary Material.

\subsection{Refinement of OPCG and M-MADE}
We can improve OPCG and M-MADE by refining the kernel weights in \eqref{eqn:opcg-loc-like} and \eqref{eqn:made-nll-loc}, similar to the refinement schemes of \cite{xia2002adaptive} and \cite{xia2007constructive}. By replacing $W_{ix}(h)$ in \eqref{eqn:opcg-loc-like} with $K\{ \hat \beta_{ \mathrm{opcg} }^{\smtop} (X_i - x)/h\}$ for OPCG, or $W_{ij}(h)$ in \eqref{eqn:made-nll-loc} with $K\{ \hat \beta_{ \mathrm{made} }^{\smtop} (X_i - X_j)/h\}$ for M-MADE, we obtain the refined objective functions for OPCG and M-MADE. 
To obtain the refined estimators, we minimize the objective functions with the fixed weights to obtain the new $\beta$. We then update the weights using this new $\beta$ and repeat the minimization. We continue this process until some convergence criterion on $\b$ is met. 
The resulting estimators are referred to as the refined OPCG estimator, denoted by $\hat \b_{ \mathrm{ropcg} }$, and the refined M-MADE estimator, denoted by $\hat \b_{ \mathrm{rmade} }$, respectively. As with the classical OPG and MAVE, the refined estimators often perform better than their unrefined counterparts.

\section{Categorical and Ordinal Categorical Responses}\label{sec:link}
	
The crux of our proposed methods {is} the multivariate link functions in our {localized} multivariate {Generalized Linear Models (GLMs)}. These links and their inverses determine the relationship between the conditional mean and the canonical parameter, through which the predictor $X$ relates to the response $Y$. 
In this section, we develop the multivariate link functions for categorical and ordinal-categorical responses. 
Special attention will be given to the canonical link, which is the Multivariate Logit link function for a categorical response, and the Multivariate Adjacent-Categories Logit link function for the ordinal-categorical response \citep{agresti2010analysis, agresti2013categorical}. All derivations are relegated to the {S}upplementary {M}aterial.

\subsection{Categorical Response}

We will only develop the canonical link in this case, as that is the most commonly used link for categorical responses.
For a given number of categories $m_0$, a categorical response $Y \in \{1,\ldots,{m_0}\}$ can be represented by $S = (S^1,\ldots,S^{ {m_0}  }) \in \{0,1\}^{ {m_0} }$. The entries of {the random vector} $S$ are $S^j=0$ for $j\neq {Y}$ and $S^{Y} = 1$. {Assuming $S$ conditioning on $X=x$ has a multinomial distribution:} $S \sim {\mbox{multinomial}(k, (p_1,\ldots,p_{ {m_0} } ) ) } $ with $k=1$, {our model is a special case of the multivariate GNM.} Since the vector $S$ and probabilities ${(p_1,\ldots,p_{ {m_0} } )}$ are constrained by $\sum_{j=1}^{ {m_0} } S^{j} = 1$ and $ \sum_{j=1}^{{m_0}  }  p_j = 1$, respectively, we set the ${m_0}$-{th} category as our baseline with $S^{ {m_0} } = 1 - \sum_{j=1}^{ {m_0} -1} S^{j}$ and $p_{ {m_0} } = 1 - \sum_{j=1}^{ {m_0} -1} p_{j}$, without loss of generality. {Let $m$ denote $m_0 - 1$.}

For convenience of notation, {reset} $S = (S^1,\ldots,S^{m})$ and $p = (p_1,\ldots,p_{m})$ {to denote} the unconstrained response and probability {vector}, respectively.
As an example, when $m_0=3$, $Y=1,2,3$ are equivalent to $S=(1,0), (0,1), (0,0)$, respectively. 
The mean and variance of $S$ are
\begin{align*}
E(S) 
= p 
, 
\quad 
{\mathrm{var}}(S)
=   {\mathrm{diag}(p)} 
- { pp^{\smtop} } 
,
\end{align*}
where, for a vector $v$, $\mathrm{diag}(v)$ denotes the diagonal matrix with $v$ as the diagonal.
{For a multinomial distribution, the canonical link and its inverse are the Multivariate Logit and Multivariate Expit transformations, defined as} 
\begin{align*}
\th = \th(p) = \log \bigg ( \frac{ p } {1 - \bs 1_m^{\smtop} p } \bigg ),
\quad 
p = p(\th) =  \frac { e^{\th } } {  1 + \bs 1_m^{\smtop} e^{\th }},
\end{align*}
respectively, { where $\theta$ is the canonical parameter and $\bs 1_m$ is a vector of ones of length $m$.}  
{T}he density and log-likelihood {of $S$} are 
\begin{align*}
f(S;p) \propto \, & \prod_{j=1}^{m} (p_j)^{S^j} \bigg( 1 - p_1 - \cdots - p_m \bigg)^{1 - S^1 - \cdots - S^m},\\ 
\ell(\th;S) = \, & \th^{\smtop}S -  \log ( 1 - \bs 1^{\smtop}_m  e^\th   ),
\end{align*}
respectively{.}  
Comparing the above log-likelihood with \eqref{eq:multivariate GLM}, we have $b(\th) =  \log (1 + \bs 1_m^{\smtop} e^{\th})$.
{Thus, the negative local log-likelihood in this case specializes to
	{\small 
		\begin{align}
		\label{eqn:opcg-nll-cat}
		\begin{split} 
		\ell_j (a_j, B_j)  
		= & \frac 1n \sum_{i=1}^{n} W_{ij}(h) ( -\{a_j + B_j^{\smtop}(X_i - X_j)\}^{\smtop} S_i  \\
		& + \log [   1 + \bs 1_m^{\smtop} \exp \{ a_j + B_j^{\smtop}(X_i - X_j) \}  ] )
		.
		\end{split} 
		\end{align}
	}\normalsize%
}

\subsection{Ordinal Response}\label{sec:ord}
  
{
	For ordinal categorical responses, we develop the multivariate canonical link and three other commonly used link functions.} 
For $m_0$ given ordered categories, an ordinal-categorical response $Y$ {is also a member of} $\{1,\ldots,m_0\}$.
{
	However, here, the numerical order of $1,\ldots,m_0$ has a practical meaning, usually representing a rank. To reflect this feature, we use another random vector, $T=(T^1,\ldots,T^{m}) \in \{0,1\}^{m}$, where $m=m_0-1$ to represent $Y$, where $T^j$ is the indicator $\i\{Y > j\}$.
} 
The vector $T$ accounts for the ordering of the categories that are not accounted for {by} $S$.
For example, with $m_0=3$ ordered categories, the ordinal responses $Y = 1, 2, 3$ are equivalent to $T = (0,0), (1,0), (1,1)$, respectively. 

Setting $T^0 \equiv 1$ and $T^{m_0} = 0$, we {obtain} a bijection between the vectors $S$ and $T$, where $S^j = T^{j-1 }- T^{j}$ for $j=1,...,m-1$. 
{We derive the distribution of $T$ via the multinomial distribution of $S$ {as follows}  
	\begin{align*}
	f({T};p) 
	\propto  & \prod_{j=1}^{m_0} (p_j)^{S^j} 
	=  
	p_1
	\bigg ( \frac{ p_2 }{p_1} \bigg )^{T^1} 
	\cdots 
	\bigg ( \frac{ p_{m_0} }{p_{m_0-1}} \bigg )^{T^{m_0-1} } 
	{.}
	\end{align*}
	Let $\g = ( \g_1, \ldots, \g_m )^{\smtop} $, with entries $\g_{j} = \sum_{k=1}^{j} p_{k}$ being the cumulative probabilities up to category $j$.  
	Note that  
	$p_{j} = \g_{j} - \g_{j-1}$, where we set 
	$\g_{0} = 0$. 
	Let $\t = {\bs 1_m}   - \g$. 
	Then the mean and variance of the random vector $T$ are
	\begin{align*}
	E(T) = \t, 
	\quad 
	{\mathrm{var}}(T) =  \Gamma - \t  \t^{\smtop} ,
	\quad \text{where} \quad 
	\Gamma   = \left (
	\begin{matrix}
	\t_{1}  & \t_{2}  & \cdots & \t_{m-1}& \t_{m} \\
	\t_{2}  & \t_{2}  & \cdots & \t_{m-1} & \t_{m} \\ 
	\vdots  & \vdots  &    & \vdots & \vdots \\ 
	\t_{m-1} & \t_{m-1} & \cdots & \t_{m-1}  & \t_{m} \\
	\t_{m} & \t_{m}  & \cdots & \t_{m} & \t_{m}
	\end{matrix}
	\right )
	.
	\end{align*}  
	{
		The distribution of $T$ belongs to an exponential family with mean parameter $\t$.
		The canonical parameter for this family is 
	}
	$\th = (\th_1,...,\th_{m} )$, where the $j$-{th} entry of $\th$ is
	\begin{align}
	\label{eqn:ad_cat_entry}
	\th_j
	= 
	\log \bigg ( \frac{ \g_{j+1} - \g_{j} }{ \g_{j} - \g_{j-1}  } \bigg )
	=
	\log \bigg ( \frac{  \t_{j} - \t_{j+1} }{  \t_{j-1} - \t_{j}  } \bigg )
	=
	\dot b (\t),
	\end{align}
	{which is the canonical link.}
	Letting
	\begin{align*}
	P = \left ( 
	\begin{matrix}
	0 & 1 \\
	I_{m} & 0 \\
	\end{matrix} \right )
	,
	\end{align*}
	we can rewrite \eqref{eqn:ad_cat_entry} in matrix notation as {$\theta = \log \{ [ \mathrm{diag} \{ ( P - I_m )\t \} ]^{-1}  ( I_m - P^{\top} )\t \}$}, which we refer to as the Adjacent-Categories link, or Ad-Cat link. 
	To derive the inverse of the Ad-Cat link, { or the mean function,} we {introduce the vector-valued function} $\phi(\th) = \{ \phi_1(\th ), \phi_2(\th ), ..., \phi_{m}(\th) \}$ as
	\begin{align*}
	\phi_{j} (\th) 
	& = \sum_{r=1}^{j} \prod_{s=1}^{r} \exp \left ( \th_{s}  \right )  
	= \sum_{r=1}^{j} \exp \left ( \sum_{s=1}^{r}  \th_{s}  \right )   
	, 
	{ \quad j=1,\ldots,m}
	.
	\end{align*}
	{In matrix notation, we have $\phi(\th) =L\exp(L \theta)$, where $L$ is a $m \times m$ matrix with its elements in the lower triangle (including the diagonal) being 1 and other elements being 0.}
	Then{, the mean function}, which maps the canonical parameter $\th$ to the mean {parameter}   $\t$, is
	\begin{align*}
	\t(\th) =
	\bigg [
	-I_{m}   
	- 
	\frac{ \{\bs 1_m  + e_1 + P \phi(\th) \} e_1^{\smtop} } { 1 - e_1^{\smtop} \{\bs 1_m  + e_1 + P \phi(\th) \}  } 
	\bigg ] 
	P \phi(\th) 
	=
	\frac{QPL\exp(L \theta) }{ 1 + e_1^{\smtop} PL\exp(L \theta) } 
	,
	\end{align*}
	where $Q = - I_{m} + \bs 1_m e_1^{\smtop} + e_1 e_1^{\smtop}$, $e_1$ being a vector of zeros except with a 1 in the first entry. The details of the derivation for the {mean function} can be found in the {S}upplementary {M}aterial. 
	{In terms of the canonical parameter}, the log-likelihood for the $T$ is 
	\begin{align}
	\label{eqn:ord-cat}
	\ell(\th; T)
	= &  
	\th^{\smtop} T - \log\{1 + e_1^{\smtop}PL\exp(L \theta) \}
	.
	\end{align}
	{ So $b(\th) = \log\{1 + e_1^{\smtop}PL\exp(L \theta) \}$.} We refer to any random vector $T$ with log-likelihood \eqref{eqn:ord-cat} as having an \textit{Ordinal-Categorical (Or-Cat)} distribution. 
	{
		If we use the canonical link, then the negative local log likelihood reduces to
		{\small  
			\begin{align}
			\label{eqn:opcg-nll-ord}
			\begin{split} 
			\ell_j (a_j, B_j)  
			= &  \frac 1n  \sum_i^{n} W_{ij}(h) \{ -\{a_j + B_j^{\smtop}(X_i - X_j)\}^{\smtop}T_i  \\
			& + \log ( 1 - e_{1}^{\smtop}  P L \exp[ L \{a_j + B_j^{\smtop} (X_i - X_j) \} ]    )   
			\} .
			\end{split} 
			\end{align}}\normalsize%
	}

		There are three other popular multivariate link functions for ordinal-categorical data.
			{
				\begin{enumerate}[1.]
			\setlength\itemsep{0.25cm}
			\item The cumulative Logit link, where $\psi(\t) = \mathrm{Logit}(\bs 1_m-\t)$, and
			\begin{align*}
			&(\t^{-1} \circ \psi ^{-1} ) (\eta) \\
			= &\log [ \mathrm{diag}\{ ( P - I_m ) \{\bs 1_m - \mathrm{expit}(\eta) \}   \}^{-1}  
			( I_m - P^{\top} ) \{\bs 1_m - \mathrm{expit}(\eta) \}  ]
			.
			\end{align*}
			
			\item The cumulative Probit link, where $\psi(\t) = \Phi^{-1} ( \bs 1_m - \t)$ and 
			\begin{align*}
			& (\t^{-1} \circ \psi ^{-1})  (\eta) \\
			= &\log [ \mathrm{diag}[ ( P - I_m ) \{\bs 1_m - \Phi(\eta) \}  ]^{-1}  
			( I_m - P^{\top} )  \{\bs 1_m - \Phi(\eta) \}  ]
			.
			\end{align*}
			
			\item The complementary log-log link, where $\psi(\t) = - \log \{ - \log ( \t ) \} $ and 
			\begin{align*}
			& (\t^{-1} \circ \psi ^{-1})  (\eta) \\
			= & \log [ \mathrm{diag}[ ( P - I_m ) \exp \{  - \exp (- \eta)  \}  ]^{-1}  
			( I_m - P^{\top} ) \exp \{  - \exp (- \eta)  \}  ]
			.
			\end{align*}
			
		\end{enumerate} }
		The sample-level objective functions for these three links are \eqref{eqn:opcg-loc-like} with $\t^{-1} \circ \psi^{-1}$ replaced by the above functions, and $b( \cdot)$ replaced by $\log\{1 + e_1^{\smtop}PL\exp(L \cdot ) \}$.

	\section{{Fisher Consistency}}\label{sec:fisher}
	
	In this section, we establish the Fisher consistency of OPCG. Here, the notion of Fisher consistency is an extension of the classical concept (see, for example, \citet[page 53]{li2019graduate}) to statistical functionals that involve a tuning parameter. Specifically, let $\mathfrak F$ be the class of all distributions of $(X,Y)$,  $F_0 \in \mathfrak F$ the true distribution of $(X,Y)$, and $F$ a generic member of $\mathfrak F$.  A \textit{tuned statistical functional} \citep[page 163]{li2018sufficient} is a function $T(F, h)$ defined on $\mathfrak F \times \R^k$, taking values in a metric space $(\mathfrak M, d)$, with $h$ being the tuning parameter. A tuned statistical functional is said to be Fisher consistent for a parameter $\theta$ if $T(F_0, h) \to \theta$ as $h \to 0$ (see Definition 11.1 of \cite{li2018sufficient}). Furthermore, suppose $\{T_a(F, h) : a \in A\} $ is a family of tuned statistical functionals. We say this family is uniformly Fisher consistent if $\sup_{a \in A} d( T_a(F,h), T_a(F_0,h) ) \to 0$ as $h \to 0$.  
	We next develop the uniform Fisher consistency of the minimizer of the objective function \eqref{eqn:exp-risk}.

	The fundamental fact underlying the Fisher consistency of our method is that, if $X$ and $Y$ satisfy the SDR relation \eqref{ass:sdr-cms} and follows the multivariate GNM \eqref{eq:fully nonparametric}, then the gradient of the canonical parameter $\theta(x)$ provides us enough information to fully recover the central mean subspace (and hence also the central subspace). The next proposition establishes this fact. For the rest of the paper, we refer to the $p \times m$ derivative matrix $\partial \theta (x)^{\smtop} / \partial x$ as the canonical gradient, and we assume the following assumption holds.  
	
	\begin{ass} 
		\label{ass:conv_u} 
		The { random vector $u(X) = \beta^{\smtop} X$ is supported on a convex set.} 
	\end{ass}
	
	\begin{prop} \label{prop:unbiased+ex}
		If $X$ and $Y$ {satisfy} \eqref{ass:sdr-cms} and {\eqref{eq:fully nonparametric}, then the following statements hold true.
		}
		\begin{enumerate}[(a)]
			\setlength\itemsep{0.25cm}
			
			\item {The columns of the canonical gradient, $\partial \th(x)^{\smtop}/\partial x \in \R^{p \times m}$, belong to $\SS_{E(Y|X)}$.
			}

			\item {If, in addition, Assumption \ref{ass:conv_u} holds, then the columns of the expected outer product of the canonical gradient span $\SS_{E(Y|X)}$; that is 
			}
			\begin{align*}
			\mathrm{span}
			\bigg [
			E \bigg \{ \frac{ \partial \theta(x)^{\smtop}  }{ \partial x} 
			\frac{ \partial \th( x)} {\partial x^{\smtop}} \bigg \} 
			\bigg ] 
			= \SS_{E(Y|X)}
			. 
			\end{align*}    
			
		\end{enumerate}
	\end{prop}
	
	{Statement (b)} implies that the $d$ {leading} eigenvectors of 
	\begin{align*}
	\Lambda = E\bigg [ \frac{\partial \th(x)^{\smtop}}{\partial x} 
	\frac{ \partial \th  (x)  }{\partial x^{\smtop}} \bigg ]
	,
	\end{align*}
	{say $\eta = (\eta^1,\ldots,\eta^d)$}, span $\SS_{E(Y|X)}$. 
	This result is analogous to OPG and motivates our approach for estimating $\SS_{E(Y|X)}$. 
	
	The statistical functional $T(F, h)$ for OPCG is the minimizer of the function
	\begin{align}
	\label{eqn:exp-risk} 
	\begin{split} 
	R_n(a, B, F_0)  & = E_F ( K\{ (X - x)/h\} [-(\t^{-1} \circ \psi^{-1}) \{a + B^{\smtop}(X-x) \}^{\smtop}Y \\
	& + ( b \circ \t^{-1} \circ \psi^{-1}) \{a + B^{\smtop}(X-x)\}] ) ,
	\end{split} 
	\end{align}
	where $E_F$ represents the integral with respect to a distribution $F$ of $(X,Y)$. 
	%
	Thus, letting $a(h, x)$ and $B(h, x)$ be the minimizer of \eqref{eqn:exp-risk} with $F = F_0$, Fisher consistency for OPCG means $a(h, x) \to \th(x)$ and $B(h, x) \to \partial \th(x)^{\smtop}/\partial x$ as $h \to 0$.
	We now make some assumptions for Fisher consistency.

	
	\begin{ass} 
		\label{ass:xy} The joint density $f(x,y)$ of $(X,Y)$ is twice continuously differentiable with respect to $x$.
		The sample space {$\om_X \subseteq \R^p$} is compact. The random vector $Y$ has finite third moments. 
	\end{ass}

	\begin{ass} 
		\label{ass:loss1} 
		The canonical parameter $\th$ is identifiable and the negative log- likelihood $\ell(\th;y) = -\th^{\smtop}y + b(\th)$ is strictly convex in $\th$, twice continuously differentiable in $\th$ and has a unique minimum. 
		{Let the derivative be denoted by $g(y, \th) = \partial \ell(\th;y)/ \partial \th $. }
		Furthermore, for each $x \in \om_X$, 
		the conditional risk 
		$R(\th,x) = 
		E\{ -\th^{\smtop}Y + b(\th) | X = x \} $ 
		is twice continuously differentiable in $\th$ with unique minimizer $\th(x)$. 
		
	\end{ass}
	\begin{ass} 
		\label{ass:par} 
		The canonical parameter $\th(\cdot): \om_X \to \Theta \subset \R^m$ is a continuously differentiable function of $x$.
		The parameter space $\Theta \subset \R^{m}$ is convex.
		The parameters $a \in \R^m$ and $B \in \R^{p \times m}$ lie in a compact and convex parameter space.  
		
	\end{ass}
	\begin{ass} 
		\label{ass:id} Derivatives and integrals in 
		\begin{align*}
		\frac{\partial}{ \partial \th} \int \int   \ell(\th;y)f(x,y) dydx
		\qquad and \qquad 
		\frac{\partial}{ \partial x} \int \frac{ \partial \ell (\th;y)}{\partial \th} f(y|x ) dy 
		\end{align*} 
		are interchangeable.
	\end{ass}  
	\begin{ass}  
		\label{ass:ker} The kernel $K$ is a symmetric {probability density function} with finite moments. Furthermore, $\int K(u) uu^{\smtop} du = I_p$ and $K(u)$ is twice continuously differentiable.
	\end{ass}
	
	\begin{ass}
		\label{ass:loss2}  
		\indent 
		
		\begin{enumerate}
			\setlength\itemsep{0.25cm}
			\item There is a compact subset $A$ of $\R^{m(p+1)}$ such that $\{c_0 (x): x \in \Omega_X \} \subseteq A$.
			\item Let $\Xi$ be the set $(0,1) \times A \times \Omega_X$ and 
			\begin{align*}
			\nu (u) =
			\begin{pmatrix}
			1 \\
			u
			\end{pmatrix} \otimes I_m
			.
			\end{align*}
			If $\lambda(h, c, x)$ is the smallest eigenvalue of the matrix
			\begin{align}\label{eq:S D inv}
			\int K (u) \nu (u) \partial_2 g ( y, \nu (h u )^{\smtop} c ) \nu (u)^{\smtop} f(x + h u, y) dy du,
			\end{align}
			then $\inf_{(h,c,x)\in \Xi} \, \lambda (h, c, x) > 0$,
			where $\partial_2 g$ denotes the partial derivative of $g$ with respect to the second argument. 
			
			\item 
			$\sup_{(h, c, x) \in \Xi} \left\|  I_1 (h, c, x )  
			+ I_2 (h, c, x ) 
			+ I_3 (h, c, x )
			+ I_4 (h, c, x )
			\right\| < \infty$,  where
			\begin{align*}
			I_{1} (h, c, x ) = & \int K(u) \nu (u) c^{\smtop} (u \otimes I_m) \partial_2^2 g(y, \nu (hu)^{\smtop} c ) (u \otimes I_m)^{\smtop} c \\
			& \times  f (x + hu, y)du dy ,   \\
			I_{2} (h, c, x ) = & \int K(u) \nu (u) \partial_2 g(y, \nu (hu)^{\smtop} c ) (u \otimes I_m)^{\smtop} c \partial_1 f (x + hu, y)^{\smtop} u du dy ,   \\
			I_{3} ( h, c, x ) = & \int K(u) g(y, \nu (hu)^{\smtop} c ) (u \otimes I_m)^{\smtop} c   \partial_1  f (x + h u, y ) u du dy, \\
			I_{4} ( h, c, x ) = & \int K(u) g(y, \nu (hu)^{\smtop} c ) u^{\smtop} \partial_1^2 f (x + h u, y ) u du dy,
			\end{align*}
			and $\| \cdot \|$ refers to operator norm.  
			
			\item For some compact set $A$, the following are finite:
			\begin{align*}
			& \sup_{ c \in A,h \in [0,1],  x \in \om_X} 
			\bigg \| \bigg \{ \int  K(u) \nu(u)  \partial_2 g(y, \nu(hu)^{\smtop} c  )   \nu(u)^{\smtop} f(x +hu, y) dydu \bigg \}^{-1}
			\bigg \| ,\\
			& \sup_{c \in A , h \in [0,1], x \in \om_X} 
			\int  K(u)  
			\bigg | g( y , \nu(hu)^{\smtop} c )^{\smtop} \nu(u)^{\smtop} \nu(u) g( y , \nu(hu)^{\smtop} c ) \bigg | \\
			& \hspace{3cm}  \times 
			f(x + hu ,y)dudy,\\
			& \sup_{c \in A , h \in [0,1], x \in \om_X} 
			\int  K(u)  
			\bigg \| \nu(u) g( y , \nu(hu)^{\smtop} c ) g( y , \nu(hu)^{\smtop} c )^{\smtop} \nu(u)^{\smtop}  \bigg \|\\
			& \hspace{3cm}  \times 
			f(x + hu, y)dudy,\\
			& 
			\sup_{c \in A , h \in [0,1], x \in \om_X} 
			\int  K(u)  
			\bigg \|
			\nu(u) \partial_2 g(y, \nu(hu)^{\smtop}   c  ) \nu(u)^{\smtop} \nu(u) \partial_2 g(y, \nu(hu)^{\smtop}   c  ) \nu(u)^{\smtop}
			\bigg \|  \\
			& \hspace{3cm}  \times f(x + hu, y)dudy 
			,
			\end{align*}
			where $\| \cdot \|$ refers to operator norm for matrices.  
		\end{enumerate}
	\end{ass}
	{
		Overall, Assumption 7 imposes conditions on the second-order pure and mixed derivatives of the loss function and joint density, in order to control the remainder of our Taylor expansions. 
		Consider the scenario where the response, $Y$, is uni-variate, i.e. where $m=1$. Assumption $7(a)$ requires the true parameter, $c_0(x)$, for all $x \in \om_X$, to lie within a compact subspace of $\R^{p+1}$. For $7(b)$, we now have $\nu (u) = (1, u)^{\smtop}$, and
		\begin{align}
		\int K (u) (1, u)^{\smtop} \partial_2 g ( y, \nu (h u )^{\smtop} c ) (1, u) f(x + h u, y) dy du,
		\end{align}
		which can be seen as the local information matrix in $c$, for given $h$ and $x$. We require that this matrix has positive smallest eigenvalue, which is a common assumption in non-parametrics. For $7(c)$, a uni-variate response implies the following mixed derivatives:
		\begin{align*}
		I_{1} (h, c, x ) = & \int K(u) (1, u)^{\smtop} c^{\smtop} u  \partial_2^2 g(y, \nu (hu)^{\smtop} c ) u^{\smtop} c f (x + hu, y)du dy ,   \\
		I_{2} (h, c, x ) = & \int K(u) (1, u)^{\smtop} \partial_2 g(y, \nu (hu)^{\smtop} c ) u^{\smtop} c \partial_1 f (x + hu, y)^{\smtop} u du dy ,   \\
		I_{3} ( h, c, x ) = & \int K(u) g(y, \nu (hu)^{\smtop} c ) u^{\smtop} c   \partial_1  f (x + h u, y ) u du dy, \\
		I_{4} ( h, c, x ) = & \int K(u) g(y, \nu (hu)^{\smtop} c ) u^{\smtop} \partial_1^2 f (x + h u, y ) u du dy.
		\end{align*}
		These mixed derivatives arise from the second-order term in the Taylor expansion in $h$ of the population-level local score function, $S(h, x, c)$. And so, assuming their supremum to be absolutely and uniformly bounded is a smoothness assumption on the local score function. 
		For $7(d)$, the first statement assumes the inverse of the local information matrix is uniformly bounded, which amounts to a smoothness assumption. The second to fourth statements assume the local score and information have finite second moments, which are necessary for the Bernstein-type inequalities we apply to our expansions. 
		
		For validating the assumptions in practice, we can check some more stringent criteria. If the function $g(y, \theta)$ is smooth in both arguments and $\Omega_Y$ is compact as well, then the assumptions hold. Similarly, if $g(y, \theta)$ and its derivatives with respect to $\theta$ are bounded uniformly over $y$ and $\theta$, then the assumptions also hold. We do not believe these smoothness assumptions are impractically stringent or more stringent than other non-parametric estimation methods.
		
		The following two theorems provide the fisher consistency of OPCG and its convergence rate. 
	}
	
	\begin{thm}
		\label{thm:fisher-consis} 
		Under Assumptions \ref{ass:xy}-\ref{ass:loss2}, the {minimizers} $a(h,x)$ and $B(h,x)$ of $R_n(a, B, F_0)$ in \eqref{eqn:exp-risk} are uniformly {F}isher consistent 
		as $h \downarrow 0$,  
		with convergence rates 
		{
			\begin{align*}
			\sup_{x \in \om_X} 
			\| a(h,x)  -   \th(x)  \|
			= O(h^2),
			\qquad 
			\sup_{x \in \om_X} 
			\bigg \| B(h,x)  - \frac{ \partial  \th(x)^{\smtop} }{\partial x}  \bigg \|  
			= O(h)
			,
			\end{align*}
		} $\!\!$where $\| \cdot \|$ refers to Euclidean norm and operator norm for matrices.  
	\end{thm}
	Furthermore, by Theorem \ref{thm:fisher-consis} and Proposition \ref{prop:unbiased+ex}, the $d$ leading eigenvectors of the expected outer product $\Lambda(h) = E\{B(h,X)B(h,X)^{\smtop}\}$, denoted by $\eta(h) \in \R^{p \times d}$, are {Fisher} consistent for the eigenvectors, $\eta$, of $\Lambda$, which span $\SS_{E(Y|X)}$.
	\begin{thm}
		\label{cor:fisher-consis-opcg} 
		Under Assumptions \ref{ass:xy}-\ref{ass:loss2}, as $h \downarrow 0$, {$\mathrm{span}\{ \eta(h)\}$} is {F}isher consistent for $\SS_{E(Y|X)}$  
		with convergence rate $\| \eta(h) \eta(h)^{\smtop} - \eta \eta^{\smtop} \|  = O(h)$.
	\end{thm}

	\section{{Consistency and Convergence Rate}}\label{sec:consis}
	
	{
		Let $\{ (\hat a(h,x), \hat B(h,x)): x \in \om_X \}$ and $\{ (\hat a_i, \hat B_i) : i = 1,\ldots,n \}$ be the OPCG estimates defined in section \ref{sec:glm_opcg}. In this section, we develop the uniform convergence rates of $\hat a(h,x)$ and $\hat B(h,x)$, as well as the convergence rate of $n^{-1} \sum_{i=1}^n \hat B_i \hat B_i^{\smtop}$. These results also imply that $\hat a(h,x)$ and $\hat B(h,x)$ are uniformly consistent, and $n^{-1} \sum_{i=1}^n \hat B_i \hat B_i^{\smtop}$ is consistent.
	}
	We make { the following} assumption on the bandwidth $h$.
	\begin{ass} 
		\label{ass7:h} There is an $s>2$ such that for all $0 < \a < \frac{s-2}{(p+4)s} $, $h \propto n^{-\a}$. 
	\end{ass} 
	{
		The next theorem asserts that $\hat a(h, x)$, $\hat B(h,x)$, for $x \in \om_X$, are uniformly consistent.
	}
	\begin{thm}
		\label{thm:consis-grad} 
		
		Under Assumptions \ref{ass:xy}-\ref{ass7:h}, for {$x \in \om_X$}, the estimates {$\hat a(h,x)$ and $\hat B(h,x)$} are uniformly consistent for the canonical parameter and its gradients at {$x$}, respectively, with convergence rates
		{
			\begin{align*}
			\sup_{x \in \om_X} \| \hat a(h,x)  - \th(x)   \|  
			& = O_{a.s}(h^2 +  \d_{ph} ),\\
			\sup_{x\in \om_X}\bigg \| \hat B(h,x)  - \frac{ \partial  \th(x)^{\smtop} }{\partial  x}  \bigg \| 
			& = O_{a.s}(h + h^{-1} \d_{ph} )
			,
			\end{align*}
		}
		where $\d_{ph} = \sqrt{ \log n / (nh^p) } \to 0$,  $h \downarrow 0$, and $h^{-1} \d_{ph} \to 0$ as $n \to \infty$ by Assumption \ref{ass7:h}.  
	\end{thm}
	
	Let $\hat \Lambda_n = n^{-1} \sum_{i=1}^n \hat B_i \hat B_i^{\smtop}$.
	The next theorem gives the convergence rate of $\hat \b_\mathrm{opcg}$. Let $\eta \in \R^{p \times d}$ be the matrix whose columns are the first $d$ eigenvectors of $\Lambda$.
	
	\begin{thm}
		\label{thm:consis-opcg} 
		Under Assumptions \ref{ass:xy}-\ref{ass7:h}, the {OPCG estimator} $\hat \b_{ \mathrm{opcg} }$ is consistent for $\eta$ with convergence rate of
		{
			\begin{align*}
			\| \hat \beta_{ \mathrm{opcg} } \hat \beta_{ \mathrm{opcg} }^{\smtop} - \eta \eta^{\smtop} \| = O_{a.s}(h + h^{-1} \d_{ph} ) 
			,
			\end{align*}
		}
		where $\d_{ph} = \sqrt{ \log n / (nh^p) } \to 0$, $h \downarrow 0$, and $h^{-1} \d_{ph} \to 0$ as $n \to \infty$ by Assumption \ref{ass7:h}.  
	\end{thm}

	\section{{Implementation} }\label{sec:imp}
	
	\subsection{Standardization}
	Before applying OPCG and MAVE, we first standardize the predictor. Specifically, let $\hat \mu$ and $\hat \Sigma$ be the sample mean and sample covariance matrix. We feed into OPCG and M-MADE the standardized predictor $\hat \Sigma^{-1/2} (X_i- \hat \mu)$, $i=1, \ldots, n$. The purpose of doing so is to make it reasonable to use a spherical kernel, such as the standard multivariate Gaussian kernel. Without standardization, we would have to use a matrix of bandwidths to account for the different scales of the components of $X$, which would significantly increase the effort for bandwidth selection.
	When $\hat \Sigma$ is ill-behaved, we marginally standardize the predictor so that all predictors have unit variance.

	\subsection{Order determination: estimating $d$} 
	
	So far, in the development of OPCG and M-MADE, we have assumed that $d$, the dimension of $\SS_{E(Y|X)}$, is known. In practice, we need to estimate $d$ in order to construct a basis for $\SS_{E(Y|X)}$. 
	An advantage of OPCG is that we can apply the recently developed order determination methods based on eigenvalues and the variation of  eigenvectors, such as the Ladle estimator or the Predictor Augmentation method \citep{luo2016combining, luo2020order}. 
	%
	%
	
	Since M-MADE estimates $\b$ directly without use of eigenvalues and eigenvectors, the Ladle plot and Predictor Augmentation methods are not applicable. 
	To estimate the dimension $d$ in the context of M-MADE, we can use a cross-validation approach similar to OPG and MAVE \cite{xia2002adaptive}. A $k$-fold cross-validation for determining $d$ is outlined for scalar $Y$ by \cite{adragni2018minimum}.

	\subsection{Bandwidth Selection}\label{sec:tune} 
	\newcommand\tr{\hbox{$\scriptscriptstyle\mathrm{(T)}$}}
	\newcommand\vt{\hbox{$\scriptscriptstyle\mathrm{(V)}$}}

	To allow the users to have the option to choose any classification method, here we propose a bandwidth selection procedure that is independent of post SDR classification. The intuition behind this approach is that a good bandwidth $h$ should separate the $m$ categories into natural clusters; that is, sufficient predictors of the same label should be close to one another, and those of different labels should be far apart. We split the sample $(X_1, Y_1), \ldots, (X_n, Y_n)$ into a training set
	$\mathbb{D}^{\tr}$ $= \{ (X_1^{\tr}, Y_1^{\tr}), \ldots, (X_{n_1}^{\tr}, Y_{n_1}^{\tr}) \}$ and a validation set 
	$\mathbb{D}^{\vt} = \{ (X_1^{\vt}, Y_1^{\vt} ), \ldots, (X_{n_2}^{\vt}, Y_{n_2}^{\vt}) \}$. 
	For a fixed $h$, we compute the SDR estimate $\hat \beta (h)$ by OPCG or M-MADE using the training set. We then construct the set of sufficient predictors in the validation set:
	\begin{align*}
	\{ \hat \beta (h)^{\smtop} X_1^{\vt}, \ldots, \hat \beta (h)^{\smtop}X_{n_2}^{\vt} \} \equiv \{ U_1, \ldots, U_{n_2} \}.  
	\end{align*}
	{ This set of sufficient predictors} is partitioned into $m$ categories according to their class labels; that is, 
	\begin{align*}
	\{ U_1, \ldots, U_{n_2} \} = \cup_{\ell = 1}^m \{ U_1^{\scriptscriptstyle(\ell)}, \ldots, U_{m_\ell}^{\scriptscriptstyle(\ell)} \} \equiv \cup_{\ell = 1}^m C_\ell. 
	\end{align*}
	Let $\bar U$ be the mean of $U_1, \ldots, U_{n_2}$, and $\bar U^{\scriptscriptstyle(\ell)}$ be the sample mean of $U_1^{\scriptscriptstyle(\ell)} , \ldots, U_{m_\ell}^{\scriptscriptstyle(\ell)}$. We then compute the within-class and between-class sum of squares (SSB) as follows
	\begin{align*}
	\mathrm{SSW}(h) = \sum_{\ell = 1}^m \sum_{k=1}^{m_\ell}( U_k^{\scriptscriptstyle(\ell)} - \bar U^{\scriptscriptstyle(\ell)} )^2, 
	\quad 
	\mathrm{SSB}(h) = \sum_{\ell = 1}^m (  \bar U^{\scriptscriptstyle(\ell)} - \bar U )^2. 
	\end{align*}
	{
	%
	The corresponding F-ratio is $SSW(h)/SSB(h)$. We repeat this procedure $r$ times. Each time we re-split the data into a training and validation set before computing the F-ratio. This results in $r$ F-ratios for each $h$, denoted by $F_1(h), F_2(h), \ldots, F_r(h)$. We then take the average of these $r$ F-ratios as our criterion, $F(h) = r^{-1} \sum_{s=1}^r F_r(h)$. We take our choice of optimal bandwidth as the minimum of this average, $h^* = \argmin_{h} r^{-1} \sum_{s=1}^r F_s(h)$.  
	}
	Our procedure to determine $h$ relies on fixing the dimension $d$ for the estimate $\hat \b(h) \in \R^{p \times d}$. We suggest setting $d=m_0-1$ since this works well in our experience.

	Sometimes it happens that there are several clusters within a class $\ell$. To accommodate these situations we use k-mean to refine the above procedure. For a predetermined  $k$, we perform k-mean clustering on each class $C_\ell$, resulting in $k$ clusters, $C_\ell^{\scriptscriptstyle(1)}, \ldots, C_\ell^{\scriptscriptstyle(k)}$, and $k$-centers $\hat \nu_\ell^{\scriptscriptstyle(1)}, \ldots, \hat \nu_\ell^{\scriptscriptstyle(k)}$. We then compute the within and between sum of squares as above for the partition $\{C_\ell^{\scriptscriptstyle(r)}: \ell = 1, \ldots, m, r = 1, \ldots, k \}$ to form the refined F-ratio $F(h)$, which is minimized to obtain the optimal $h$. 
	We can further fine-tune this procedure by allowing $k$ to vary by label, but in our experience, a fixed pre-determined $k$ across all labels works well. 
	We set $k=1$ as our default. To determine whether to set $k>1$, we suggest plotting the sufficient predictors on the validation set using an inverse regression method, such as Directional Regression. This visualization provides a reasonable guide for whether the class labels have multiple clusters.


	\section{Simulations and Data {Applications }}\label{sec:examples}
	
	Our main focus will be on the performance of OPCG in simulations and applications. 
	Under our assumptions, 
	OPCG has a unique gradient estimate at each $X_i$ and can be computed in parallel to mitigate the computational cost.  
	Meanwhile, M-MADE will only be studied in our simulations due to the high computational cost, for even moderate $p$, of Step 2 and Step 3 in Algorithm \ref{alg:made}.
	{ From our simulations, we will see that M-MADE performs better than, but is still comparable, to OPCG,} similar to observations made by \cite{xia2006asymptotic} and \cite{wang2008sliced}.
	
	\subsection{Simulation Study}
	{ We conduct three simulation studies, where two have categorical responses and one has an ordinal response.
	For the first simulation,} we generate $X=(X^1,\ldots,X^{10}) \in \R^{10}$ as follows. The entries $(X^3, X^7)$ are from one of five bivariate normal distributions: 
	\begin{align*}
	(X^3, X^7) \sim N_2(\mu_c, 0.25 \times I_2), 
	\qquad \mu_c \in \{ (0,0), (3,3), (-3,-3), (-2,2), (2,-2) \},  
	\end{align*}
	while the remaining entries are generated from standard normals. The choice of $X^3$ and $X^7$ is so our initial values for M-MADE do not span the central mean subspace. 
	The five clusters determined by $\mu_c$ are labeled into three classes: samples with $\mu_c \in \{ (-2,2), (2,-2)\}$ are class `1', samples with $\mu_c \in  \{(3,3), (-3,-3)\}$ are class `2', and samples with $\mu_c = (0,0)$ are class `3'.
	The categorical response is $Y \in \{1,2,3\}$ with $m_0=3$ categories. The central mean subspace $\SS_{E(Y|X)}$  is spanned by 
	\begin{align*}
	\b = ( ( 0,0,1,0,0,0,0,0,0,0 )^{\smtop}, ( 0,0,0,0,0,0,1,0,0,0 )^{\smtop} )  \in \R^{10 \times 2}
	.
	\end{align*}
	In one simulation run, for each $\mu_c$, we sample $80$ observations for our training set, and $30$ for our testing set, giving us a total of $400$ and $150$ observations for training and testing, respectively. We consider only one simulation run for the figures and illustrations in this section. 
	
	We now use one simulation run to illustrate the tuning process.
	For tuning the bandwidth, the 5-fold supervised k-means tuning procedure outlined in section \ref{sec:imp} is applied to the training set. The 5-fold F-ratio is illustrated in Figure \ref{fig:sim_km} along with the F-ratio for one specific fold. The sufficient predictors constructed on the validation set for various values of $h$ for the specific fold are given in Figure \ref{fig:sim_tune_plots}. We set the pre-specified number of clusters per class set to $c_l=2$ and fix the dimension of the OPCG estimates to $d=2$. 
	The 5-fold cross-validation suggests a bandwidth of $h \approx 1.26$.
	\begin{figure}[!h]
		\centering
		\subfloat[Sufficient Predictors for different bandwidth, $h$]{{
				\label{fig:sim_tune_plots}
				\includegraphics[width=1\linewidth]{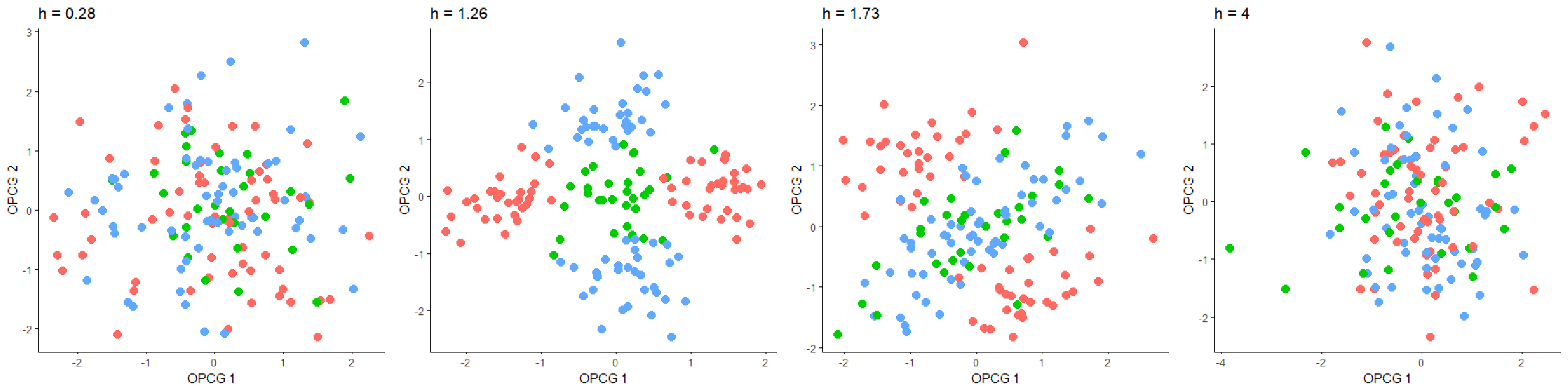}
		}}%
		\qquad
		\subfloat[K-means Tuning Procedures for $h$]{{ 
				\label{fig:sim_km}
				\includegraphics[width=.5\linewidth]{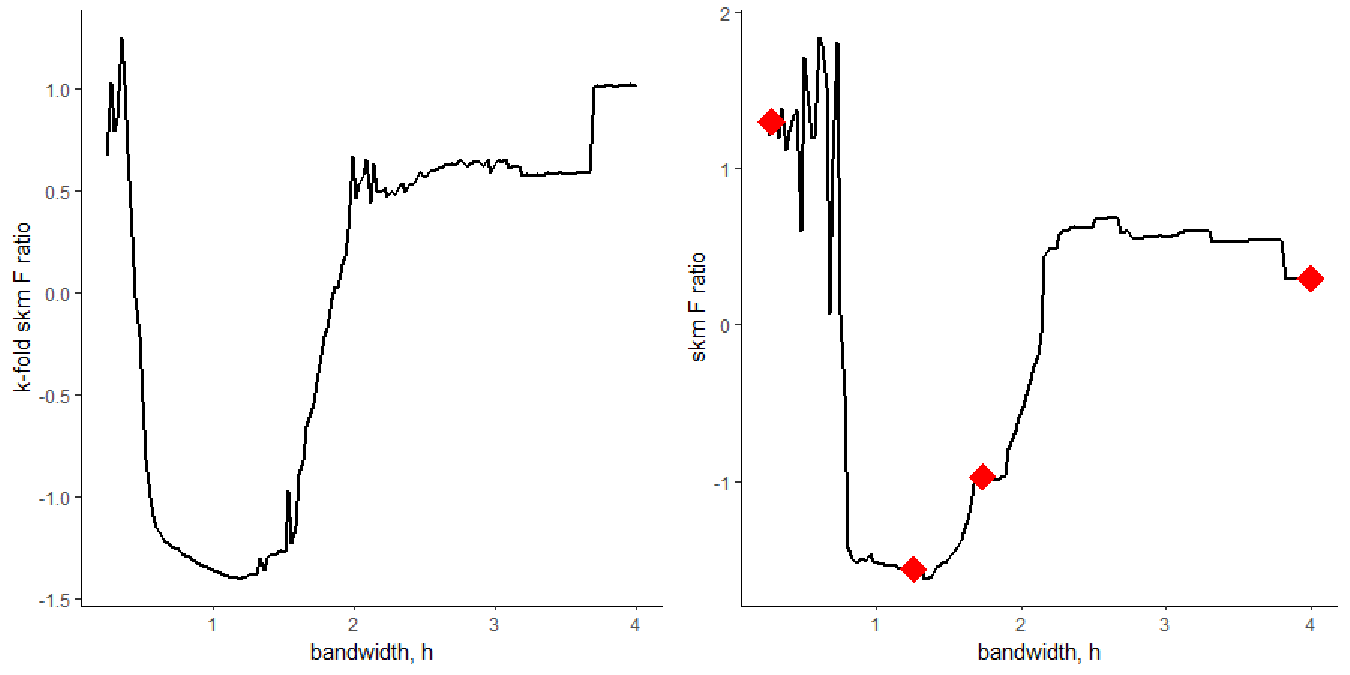}
		}}
		\caption{(a) From Left to Right: The sufficient predictors constructed on the validation set, for $h=0.28, 1.26, 1.73, 4$. Red is class $1$; blue is 2; and green is 3.
			(b) The F-ratios for 5-fold supervised k-means as applied to the training set (left) and a specific validation set (right). The four red dots in the right plot correspond to the instances of $h=0.28, 1.26, 1.73, 4$ plotted in figure (a). }\label{fig:sim_tune}
	\end{figure}
	To demonstrate the usefulness of allowing multiple clusters per class, we provide the tuning plots for $c_l=1,2,3$ in the Supplementary Material, section 7. These plots show how under specifying the number of clusters per class is detrimental to the F-ratio, whereas over-specifying the number of clusters per class appears less harmful. This reinforces the need to determine whether multiple clusters per class exist using inverse regression methods like Directional Regression.   
	For order determination, we use the predictor augmentation method of \cite{luo2020order} to determine the order for OPCG with bandwidth set at $h=1.26$. We append $p/5$ noisy predictors, as suggested by \cite{luo2020order}, for 200 replications.      
	In our experience, the bandwidth does not need to be optimized for ladle or predictor augmentation methods to work well. The Predictor Augmentation plot is illustrated in Figure S2 in the Supplementary Material; the method correctly estimates the dimension as $d=2$.

	For OPCG, we implement Algorithms \ref{alg:opcg} and \ref{alg:opcg_newton}. For M-MADE, we implement Algorithm \ref{alg:made} using Fisher-Scoring.  
	We compare OPCG with OPG, Sliced Inverse Regression (SIR; \citet{li1991sliced}) and Directional Regression  (DR; \citet{li2007directional}). 
	For the initial values of $\beta$ in M-MADE, we use $(e_1,e_2)$, where $e_i \in \R^p$ has $1$ in the $i^{th}$ position and $0$ elsewhere.  
	For dealing with multi-label classification, \cite{adragni2018minimum} suggested considering all pairwise binary classifications, while \cite{lambert2006local} suggested considering a binary classification per-label. 
	We implement these pairwise and per-label approaches with some modifications on how to select the final SDR directions.
	For the pairwise method, we use OPCG to estimate two SDR directions for each pair of classes, $\{1,2\}$, $\{1,3\}$ and $\{2,3\}$, giving us $6$ directions in total. We then take the average outer product of these 6 directions and use the two leading eigenvectors as the Pairwise method (PW-method) estimator.  
	For the per-label method, we use OPCG to estimate two SDR directions using binary logistic regression for each class in $\{1,2,3\}$, giving us $6$ directions in total. We then take the average outer product of these 6 directions and use the two leading eigenvectors as the Per-Label method (PL-method) estimator.
	We also compare OPCG with GSIM \citep{lambert2006local} from the `plsgenomics' package in R. GSIM always returns $m_0-1$ SDR directions, which happens to be the true dimension of the central mean subspace. 
	
	Our comparison metric is the average distance in Frobenius norm of the subspace spanned by the estimated SDR directions, $\SS(\hat \b)$, to the true subspace $\SS(\b)$. The distance between two subspaces spanned by matrices $A$ and $B$ is computed using the difference between their projection matrices: 
	\begin{align*}
	d(\mathrm{span} (A), \mathrm{span} (B)) = \| A(A^{\smtop}A)^{-1}A^{\smtop} - B(B^{\smtop} B)^{-1}B^{\smtop} \|_{\mathrm{F}}.
	\end{align*}
	
	The average is taken over { 100 simulation runs and the results and their standard deviations} are reported in Table \ref{tab:sim_dist}. From Table \ref{tab:sim_dist}, {SIR performs poorly because some clusters of $X$ within each category of $Y$ have symmetric support, which is a known drawback of SIR} \citep{cook1991comment}. 
	Similarly, GSIM fails due to the symmetry of the clusters since GSIM is analogous to ADE and suffers the same drawbacks as the ADE estimator \citep{xia2002adaptive}. 
	The pairwise and per-label estimates perform better than OPG, but worse than OPCG and M-MADE, demonstrating the benefit of using a multivariate link function in our approach. 
	OPCG and M-MADE outperforms every method except DR. 
	{
		We believe Directional Regression performs well here because the clusters are generated using a bivariate Gaussian and the variance of each class is oriented in different directions. From our experience, this set up is usually favorable to Directional Regression in practice.
	}
	For M-MADE, the non-uniqueness of $\beta$ resulted in slower convergence. While the convergence criterion was not met, M-MADE still performed second best in terms of distance to the true subspace. 
	Directional Regression performed the best among all methods compared in our simulations.
%
%
	%
	We provide a plot of the sufficient predictors constructed on the test set for one simulation run in section 7 of the Supplementary Materials to support the results in Table \ref{tab:sim_dist}.
	
	{
	\begin{table}[!htp]
		\caption{\label{tab:sim_dist} Mean Distances $|| \hat \b(\hat \b^{\top}\hat \b)^{-1} \hat \b^{\top} - \b(\b^{\top}\b)^{-1}\b^{\top}||_\mathrm{F}$ over 100 repetitions. Standard deviations are given in parentheses.} 
		\centering 
		\hspace*{0cm}\resizebox{\columnwidth}{!} { %
			\fbox{
				\begin{tabular}{*{8}{c}}   
					OPCG & M-MADE  & OPG & PW-method & PL-method & GSIM & DR & SIR  \\ \hline
					0.376 (0.061) & 0.282 (0.059) & 0.724 (0.137) & 0.395 (0.093) & 0.399 (0.083) & NA (NA) & 0.478 (0.089) & 0.95 (0.12)
				\end{tabular} 
		}	 }
	\end{table}
	}
	
	


	{ 
		For the second simulation, we construct bivariate gaussian clusters as in the first example. But now we have 15 clusters of size 65, whose means are dispersed in a more complicated manner. The categorical response is still 3 classes, with 4 clusters assigned to class `1' and class `3' each. The remaining 7 clusters are assigned to class `2'. For conciseness, we refer to the second plot in Figure \ref{fig:sim2_tune_plots} for an illustration of how the 15 standardized clusters are dispersed in the central subspace. 
		The central subspace is spanned by $(e_1, e_2)$, the natural euclidean basis for $\R^2$. 
		The bivariate gaussian predictors are augmented with 8 standard normals for noise. 
		We use the 5-fold supervised k-means tuning procedure to determine the bandwidth. 
		For tuning, the data are partitioned into a training set of 15 clusters or size 50, and a validation set of 15 clusters of size 15.
		For a given fold, the sufficient predictors constructed on the validation set for various values of $h$ is given in Figure \ref{fig:sim2_tune_plots}. 
		We set the pre-specified number of clusters per class to $c_l=5$, as suggested by plots of the F-ratio criterion computed using $c_l=1,5,10$, given in section 7 of the Supplementary Materials. 
		The 5-fold cross-validation suggests a bandwidth of $h \approx 0.80$.
		We fix the dimension of the OPCG estimates to $d=2$, treating it as known. The Ladle and Predictor Augmentation plots correctly select $d=2$, but are omitted for brevity.  
		\begin{figure}[!h]
			\centering
			\subfloat[Sufficient Predictors for different bandwidth, $h$]{{
					\label{fig:sim2_tune_plots}
					\includegraphics[width=1\linewidth]{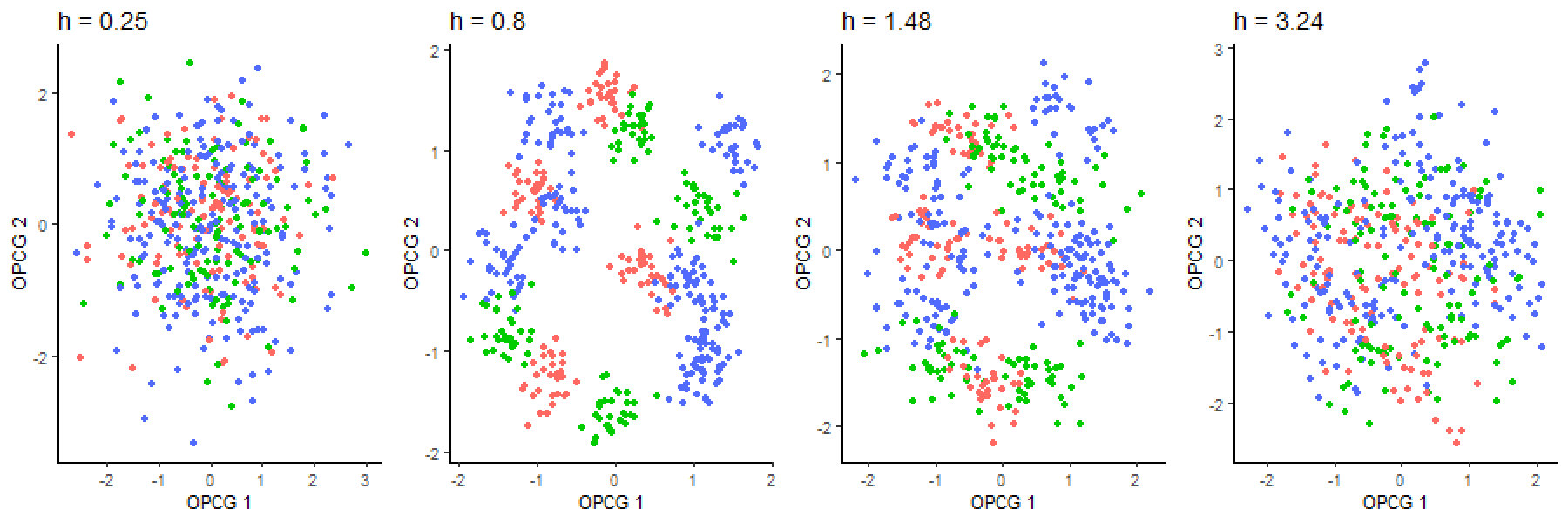}
			}}%
			\caption{The sufficient predictors constructed on the validation set, for $h=0.25, 0.80, 1.48, 3.24$. Red is class $1$; blue is 2; and green is 3. }\label{fig:sim2_tune}
		\end{figure} 
	
		The average Frobenius norm over 100 repetitions is reported in Table \ref{tab:sim_dist2}. We were unable to implement GSIM for this simulation due to computational resource limits. The results for this simulation demonstrate the effectiveness of OPCG and M-MADE over inverse regression methods when the relationship between repsonse and predictor is more complex. The pairwise and per-label approaches outperform DR in this setting as well.  
		\begin{table}[!htp]
			\caption{\label{tab:sim_dist2} Mean Distances $|| \hat \b(\hat \b^{\top}\hat \b)^{-1} \hat \b^{\top} - \b(\b^{\top}\b)^{-1}\b^{\top}||_\mathrm{F}$ over 100 repetitions. Standard deviations are given in parentheses.} 
			\centering 
			\hspace*{0cm}\resizebox{\columnwidth}{!} { %
				\fbox{
					\begin{tabular}{*{8}{c}}   
						OPCG & M-MADE  & OPG & PW-method & PL-method & GSIM & DR & SIR  \\ \hline
						0.376 (0.061) & 0.282 (0.059) & 0.724 (0.137) & 0.395 (0.093) & 0.399 (0.083) & NA (NA) & 0.478 (0.089) & 0.95 (0.12)
					\end{tabular} 
			}	 }
		\end{table}
	
	For an ordinal response example, we modify the previous categorical simulation. The predictors are 15 clusters of bivariate gaussians that are augmented with 8 standard normal as noise. 
	In this ordinal simulation, the response depends on the first coordinate of the cluster mean. Clusters with means where the first coordinate is less than -2, are assigned to class `1'. Clusters with means whose first coordinate is between -2 and 2 are assigned class `2'. And the remaining clusters are class `3'.   
	For conciseness, we refer to the first plot in Figure \ref{fig:sim2_tune_plots} for an illustration of how the 15 standardized clusters are dispersed in the central subspace. 
	The central subspace is now spanned by one direction, $e_1$. 
	To examine the sensitivity of our method to the link specification, $\psi$, in the Multivariate Generalized Nonlinear Model \eqref{eqn:noncan density}, we implement OPCG with a cumulative logistic link as well. 
	We use a bandwidth of $h=0.75$ out of convenience and treat the dimension of central subspace as known, $d=2$. A plot of the estimated sufficient predictors is provided in Figure \ref{fig:sim3_ord_plots}. The average Frobenius norm over 100 simulation runs is provided in Table \ref{tab:sim_dist3}. 
	
	\begin{figure}[!ht]
		\centering
			\subfloat{{
					\includegraphics[width=\linewidth]{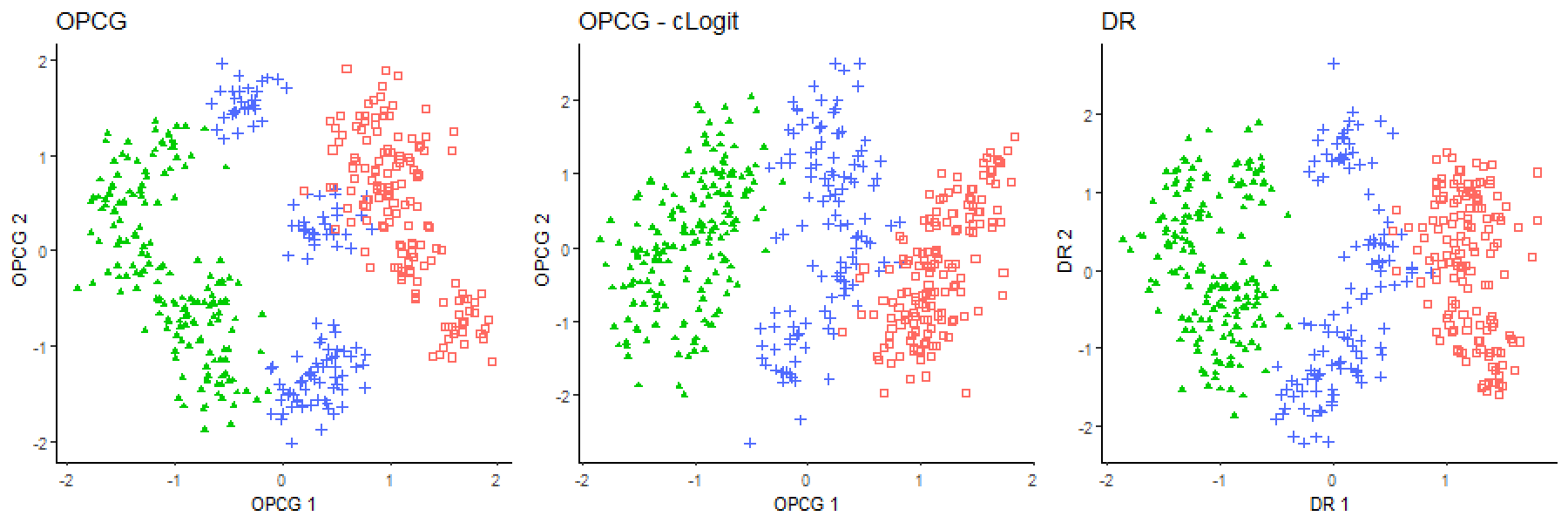}
			}}
		\caption{The sufficient predictors constructed for OPCG, OPCG with cumulative logit link, and Directional Regression. Red is class $1$; blue is 2; and green is 3. }\label{fig:sim3_ord_plots}
	\end{figure}
 	
 	\begin{table}[!htp]
 		\caption{\label{tab:sim_dist3} Mean Distances $|| \hat \b(\hat \b^{\top}\hat \b)^{-1} \hat \b^{\top} - \b(\b^{\top}\b)^{-1}\b^{\top}||_\mathrm{F}$ over 100 repetitions for ordinal simulation. Standard deviations are given in parentheses.} 
 		\centering 
 		\hspace*{0cm}\resizebox{\columnwidth}{!} { %
 			\fbox{
 				\begin{tabular}{*{8}{c}}   
 					OPCG & OPCG - clogit  & OPG & PW-method & PL-method & DR & SIR  \\ \hline
 					0.135 (0.034) & 0.178 (0. 042) & 0.252 (0.014) & 0.242 (0.072) & 0.233 (0.106) & 0.269 (0.006) & 0.270 (0.005)
 				\end{tabular} 
 		}	 }
 	\end{table}
 	From Table \ref{tab:sim_dist3}, we observe that OPCG with the canonical ad-cat link performs the best, followed by OPCG with the cumulative logistic link. In particular, both OPCG implementations perform better than the inverse regression approaches. This demonstrates OPCG can work well under link mis-specification.
	}

	\subsection{Data Applications}
	
	For our real data applications, we compare OPCG to OPG, SIR and DR by assessing the post dimension reduction classification errors using a Support Vector Machine (SVM) from the R library `e1072'. 
	For each SDR method, we train an SVM on the dimension reduced sufficient predictors constructed from the training set and compute the classification error on the sufficient predictors from the testing set. 
	In all examples, we report the classification errors for various dimensions and indicate the estimate dimension when possible. 
	
	To improve the computational speed of OPCG, we use the hybrid conjugate gradient (hCG) algorithm of \cite{dai2001efficient} to minimize the full negative log-likelihoods, which involve equations \eqref{eqn:opcg-loc-like} and \eqref{eqn:made-nll-loc}, in Step 1 of Algorithm \ref{alg:opcg}.
	An Armijo backtracking rule is used to determine the step size for the hCG algorithm \citep[page 69]{bertsekas2015convex}. 
	To further improve the computational speed of OPCG, we enforce the Armijo-Goldstein bounds for conjugate gradients \citep{nesterov2003introductory} instead of the weak Wolfe conditions used by \cite{dai2001efficient}. Without the weak Wolfe conditions, global convergence of the hCG algorithm is not guaranteed. But from our experience, the initial value from the canonical link is close enough to the minimum so that the Armijo-Goldstein bounds suffice.

	\subsubsection{Categorical Data Applications}
	We analyze three datasets with a categorical response: the pendigits and ISOLET datasets from the UCI Machine Learning repository, and the USPS handwritten digits dataset \citep{uspsdata}. 
	In pendigits and USPS, the response is a handwritten digit from 0 to 9. For ISOLET, the response is a spoken letter from ``a" to ``z". The $16$ predictors in pendigits represent features of a processed image of a digit. The $256$ predictors in USPS represent the vectorized $16 \times 16$ pixel image of a handwritten digit. The $617$ predictors in the ISOLET dataset are features extracted from the spoken alphabetic character; additional details can be found in \citep{isoletdata}. Furthermore, the ISOLET dataset is split into 5 groups of speakers, designated ISOLET-1 through ISOLET-5.
	For pendigits and USPS, we report the average classification error across 20 replications, where we re-sample the training and testing set for each replication run. We report the classification error for only one run of ISOLET because of computational costs.  
	
	For the pendigits dataset, we sample 1000 observations from the designated training set and 1000 from the designated testing set. For the USPS dataset, we only work with the 2007 observations from the designated testing set, and randomly sample 1000 for training and use the remaining 1007 for testing. 
	And for the ISOLET dataset, we use all 6238 observations in the designated training set, which is comprised of ISOLET-1 though ISOLET-4. And we use all 1554 observations in the designated testing set, which is ISOLET-5. 
	In pendigits, we standardize the predictors using the sample covariance matrix for OPCG and OPG. 
	%
	For USPS and ISOLET, we standardize the predictors marginally for OPCG and OPG, so that the predictors have unit variance. 
	%
	%
	
	We applied the Predictor Augmentation method \citep{luo2020order} to one replication run for pendigits and USPS, using $p/5$ and $p/10$ augmented noise variables, respectively. This produced an estimated dimension of $9$ for the central mean subspace in pendigits and USPS. The choices of $p/5$ and $p/10$ performed well in \cite{luo2020order}, but for our application to USPS, $p/5$ augmented variables resulted in an estimated dimension of $3$, which may be too small. The Predictor Augmentation plots are shown in Figure S4 in the Supplementary Materials. 
	Due to limited computational resources, we do not estimate the dimension for the ISOLET dataset.  
	For a single replication run of pendigits and USPS, we use our 5-fold supervised k-means tuning procedure, with $d$ set to $m_0-1=9$. For the number of clusters per class, we use $3$ in pendigits and $1$ in USPS. This was roughly determined by visualizing the sufficient predictors using Directional Regression. 
	The 5-fold supervised k-means tuning method suggests a bandwidth of $h \approx 1.42$ for pendigits and $h \approx 6.48$ for USPS.
	%
	To determine the bandwidth for ISOLET, we used ISOLET-1 for estimation and ISOLET-2 for validation. We then re-estimated OPCG on the entire training set. Applying the supervised k-means procedure on ISOLET-2, with $d=m_0-1=25$ dimensions, and one cluster per class, the suggested bandwidth is $h \approx 2.39$. 
	Plots of the supervised k-means F-ratio for pendigits, USPS, and ISOLET are provided in Figure S5 in the Supplementary Materials. 
	When comparing OPCG with SIR and DR in the ISOLET application, we use a Tikhonov regularized covariance for standardizing $X$ in SIR and DR, similar to \cite{zhong2005rsir}. The regularization parameter was roughly chosen to be $3$ from an elbow plot of classification error. 
	%
	The classification errors are reported in Table \ref{tab:cat_error}.

	\begin{table}
		\caption{\label{tab:cat_error} \% Classification Error for USPS and ISOLET. 
			The reported error rates for pendigits and USPS are averages from 20 replications.
			The ISOLET error rate is only for one replication. 
			The NA for SIR is because only $m_0-1$ SDR directions can be recovered. 
			The $^\dagger$ indicates the estimated dimension $d$ by Predictor Augmentation.  
		} 
		\centering   
		\fbox{  
			\begin{tabular}[!h]{*{6}{c}} 
				Dataset               	      & $d$ & OPCG & OPG  & DR   & SIR  \\  \hline 
				& 3 &  21.17 & 28.90  & 35.09 & 32.21  \\  
				Pendigits &  6 &   7.41 & 7.92 & 13.83 &  9.90    \\
				$p=16$, $h=1.42$ & $9^\dagger$  & 2.03 & 3.43 &  7.36 & 4.63    \\
				& 11  & 1.61 &  2.13 &  4.48 &  NA     \\
				& 13  &  1.45 &  1.61  &  2.98 &  NA     \\ \hline 
				& 3 & 36.51 &  57.73 &  39.50 & 36.13  \\ 
				& 5   & 21.55 & 53.88 & 28.43 & 23.38     \\ 
				& 7   & 14.17 & 50.83 & 23.94 & 18.00  \\ 
				USPS & $9^\dagger$   & 11.89 & 49.55 & 20.47 & 16.96    \\ 
				$p=256$, $h=6.48$ & 13  & 11.56 & 47.46 & 18.79 &  NA      \\
				& 20  & 11.00 & 44.42 & 19.66   &  NA   \\
				& 25  & 10.75 & 42.77 & 20.46  &  NA     \\ \hline 
				%
				& 5   & 50.80 & 54.27&  27.65 & 25.79  \\ 
				& 10  & 20.72 & 32.71 & 13.34 & 12.19   \\ 
				ISOLET & 15  & 14.95 &21.17& 10.58 & 10.26  \\ 
				$p=618$, $h=2.39$ & 20  & 10.97 &15.33 & 8.92  & 9.43    \\
				& 25  &  6.29 & 11.80 &  7.31 & 9.17	      \\ 
				& 30  &  6.29 & 10.65 & 6.48  	& NA    \\ 
			\end{tabular} 
		}%
	\end{table}  
	
	
	\subsubsection{Ordinal-Categorical Data Application}
	
	For ordinal-categorical responses, we analyze the red wine quality data \citep{winedata} from the UCI repository. 
	The response is a sensory-based integer score between $0$ and $10$, which is naturally ordinal. We combine scores so that the ordinal categories are $\{3,4,5\}$, $\{6\}$ and $\{7, 8\}$, since the relatively extreme scores had fewer or no observations. 
	We report the average classification error from 20 replication runs.
	For each replication run, we randomly sample two-thirds for training and use the remaining third for testing. 
	
	\begin{table}
		\caption{\label{tab:ord_error} \% Classification Error for Ordinal-Categorical Responses. The reported error rates are averages from 20 replications. } 
		\centering   
		\fbox{ 
			\begin{tabular}{*{6}{c}} 
				Dataset & $d$ & OPCG & OPG & DR & SIR \\  \hline 
				Wine & 1 & 36.40 & 36.77& 44.09& 34.90     \\ 
				$p=11$, $h=2.47$ & 2 & 36.02 &36.59& 42.40& 35.83    \\ 
				%
			\end{tabular} 
		}
	\end{table}  
	
	We use a 5-fold k-means tuning procedure, with the dimension of the sufficient predictors set to $d=m_0-1=2$, and pre-specify the number of clusters per class to be 1. This suggests a bandwidth of $h \approx 2.47$. A plot of the supervised k-means F-ratio is provided in Figure S6 in the Supplementary Materials.
	We conduct a simple ordinal classification by considering two binary classification problems for the three ordinal classes using SVM, as in \cite{waegeman2009ensemble}, which we refer to as multi-class ordinal SVM (MCOSVM).
	We train and predict using the MCOSVM as we did for the categorical analysis. 
	Instead of estimating the dimension $d$, we report the MCOSVM classification errors for dimensions $d=1$ and $d=2$ in Table \ref{tab:ord_error}.
	To supplement the ordinal classification errors, we also provide plots for the sufficient predictors constructed on the testing set in Figure S7 in the Supplementary Materials.

	\section{Conclusion}\label{sec:con}

	By imposing multivariate link functions on the conditional mean, we generalize OPG and MAVE to the Outer Product of Canonical Gradients (OPCG) and the Multivariate Minimum Average Deviance Estimator (M-MADE), which can handle categorical and ordinal-categorical responses effectively.
	%
	%
	For ordinal-categorical responses, we derived an associated Or-Cat random vector, which has the Ad-Cat and inverse Ad-Cat link functions. We showed these links are canonical and that Or-Cat random vectors are a linear exponential family, meaning OPCG is applicable.
	The OPCG estimator can recover the central mean subspace exhaustively and consistently under some assumptions.
	We also introduce a supervised k-means tuning procedure for determining the bandwidth for OPCG (and M-MADE) that performs reasonably well in our simulations and applications.  
	Our simulations and data analyses demonstrate the overall improvement of OPCG over OPG, especially in higher dimensions.
	The simulations also demonstrate OPCG's improved efficiency from using a multivariate link instead of the pairwise or per-label suggestions in existing uni-variate extensions of OPG and MAVE.
	The results from our applications indicate that OPCG is comparable, and can outperform popular inverse regression methods such as SIR and DR.  
	These multivariate link functions may be applied to other SDR methods, such as the gradient-based Kernel Dimension Reduction of \cite{fukumizu2014gradient}.

	\label{Bibliography}

\begin{acks}[Acknowledgments]
	We would like to thank two referees for their helpful suggestions and comments. Bing Li's research is supported in part by the National Science Foundation grant DMS-2210775.
\end{acks}

\begin{supplement}
All proofs and derivations are relegated to an online Supplementary Materials. This supplement also contains additional supporting figures.
\end{supplement}

\bibliographystyle{imsart-nameyear} 
\bibliography{bib_sdr,bib_opg} 


\end{document}